\documentclass[ALICE,manyauthors]{cernphprep}
\usepackage[comma,square,numbers,sort&compress]{natbib}
\usepackage{hyperref}
\usepackage{lineno}
\usepackage{xspace}
\usepackage{subcaption}
\usepackage{multirow}
\usepackage{graphicx}
\usepackage[dvipsnames]{xcolor}
\usepackage{upgreek}
\usepackage[T1]{fontenc}

\begin{document}
%

\newcommand{\pp}           {pp\xspace}
\newcommand{\ppbar}        {\mbox{$\mathrm {p\overline{p}}$}\xspace}
\newcommand{\XeXe}         {\mbox{Xe--Xe}\xspace}
\newcommand{\PbPb}         {\mbox{Pb--Pb}\xspace}
\newcommand{\pA}           {\mbox{pA}\xspace}
\newcommand{\pPb}          {\mbox{p--Pb}\xspace}
\newcommand{\AuAu}         {\mbox{Au--Au}\xspace}
\newcommand{\dAu}          {\mbox{d--Au}\xspace}

\newcommand{\s}            {\ensuremath{\sqrt{s}}\xspace}
\newcommand{\snn}          {\ensuremath{\sqrt{s_{\mathrm{NN}}}}\xspace}
\newcommand{\pt}           {\ensuremath{p_{\rm T}}\xspace}
\newcommand{\pT}           {\ensuremath{p_\mathrm{T}}\xspace}
\newcommand{\pTchjet}      {\ensuremath{p_\mathrm{T,ch\,jet}}\xspace}
\newcommand{\kT}           {\ensuremath{k_\mathrm{T}}\xspace}
\newcommand{\sqsn}         {\ensuremath{\sqrt{s_\mathrm{NN}}}\xspace}
\newcommand{\sqs}          {\ensuremath{\sqrt{s}}\xspace}

\newcommand{\meanpt}       {$\langle p_{\mathrm{T}}\rangle$\xspace}
\newcommand{\ycms}         {\ensuremath{y_{\rm CMS}}\xspace}
\newcommand{\ylab}         {\ensuremath{y_{\rm lab}}\xspace}
\newcommand{\etarange}[1]  {\mbox{$\left | \eta \right |~<~#1$}}
\newcommand{\yrange}[1]    {\mbox{$\left | y \right |~<~#1$}}
\newcommand{\dndy}         {\ensuremath{\mathrm{d}N_\mathrm{ch}/\mathrm{d}y}\xspace}
\newcommand{\dndeta}       {\ensuremath{\mathrm{d}N_\mathrm{ch}/\mathrm{d}\eta}\xspace}
\newcommand{\dndptBjet}    {\ensuremath{\mathrm{d}N_\mathrm{b-jet}/\mathrm{d}\pt}\xspace}
\newcommand{\avdndeta}     {\ensuremath{\langle\dndeta\rangle}\xspace}
\newcommand{\dNdy}         {\ensuremath{\mathrm{d}N_\mathrm{ch}/\mathrm{d}y}\xspace}
\newcommand{\Npart}        {\ensuremath{N_\mathrm{part}}\xspace}
\newcommand{\Ncoll}        {\ensuremath{N_\mathrm{coll}}\xspace}
\newcommand{\dEdx}         {\ensuremath{\textrm{d}E/\textrm{d}x}\xspace}
\newcommand{\RpPb}         {\ensuremath{R_{\rm pPb}}\xspace}
\newcommand{\RpPbBjet}     {\ensuremath{R_{\rm pPb}^\text{b-jet}}\xspace}
\newcommand{\RAA}          {\ensuremath{R_{\rm AA}}\xspace}

\newcommand{\nineH}        {$\sqrt{s}~=~0.9$~Te\kern-.1emV\xspace}
\newcommand{\seven}        {$\sqrt{s}~=~7$~Te\kern-.1emV\xspace}
\newcommand{\twoH}         {$\sqrt{s}~=~0.2$~Te\kern-.1emV\xspace}
\newcommand{\twosevensix}  {$\sqrt{s}~=~2.76$~Te\kern-.1emV\xspace}
\newcommand{\five}         {$\sqrt{s}~=~5.02$~Te\kern-.1emV\xspace}
\newcommand{\twosevensixnn}{$\sqrt{s_{\mathrm{NN}}}~=~2.76$~Te\kern-.1emV\xspace}
\newcommand{\fivenn}       {$\sqrt{s_{\mathrm{NN}}}~=~5.02$~Te\kern-.1emV\xspace}
\newcommand{\LT}           {L{\'e}vy-Tsallis\xspace}
\newcommand{\GeVc}         {Ge\kern-.1emV/$c$\xspace}
\newcommand{\MeVc}         {Me\kern-.1emV/$c$\xspace}
\newcommand{\TeV}          {Te\kern-.1emV\xspace}
\newcommand{\GeV}          {Ge\kern-.1emV\xspace}
\newcommand{\MeV}          {Me\kern-.1emV\xspace}
\newcommand{\GeVmass}      {Ge\kern-.2emV/$c^2$\xspace}
\newcommand{\MeVmass}      {Me\kern-.2emV/$c^2$\xspace}
\newcommand{\lumi}         {\ensuremath{\mathcal{L}}\xspace}

\newcommand{\ITS}          {\rm{ITS}\xspace}
\newcommand{\TOF}          {\rm{TOF}\xspace}
\newcommand{\ZDC}          {\rm{ZDC}\xspace}
\newcommand{\ZDCs}         {\rm{ZDCs}\xspace}
\newcommand{\ZNA}          {\rm{ZNA}\xspace}
\newcommand{\ZNC}          {\rm{ZNC}\xspace}
\newcommand{\SPD}          {\rm{SPD}\xspace}
\newcommand{\SDD}          {\rm{SDD}\xspace}
\newcommand{\SSD}          {\rm{SSD}\xspace}
\newcommand{\TPC}          {\rm{TPC}\xspace}
\newcommand{\TRD}          {\rm{TRD}\xspace}
\newcommand{\VZERO}        {\rm{V0}\xspace}
\newcommand{\VZEROA}       {\rm{V0A}\xspace}
\newcommand{\VZEROC}       {\rm{V0C}\xspace}
\newcommand{\Vdecay} 	   {\ensuremath{V^{0}}\xspace}

\newcommand{\ee}           {\ensuremath{e^{+}e^{-}}} 
\newcommand{\pip}          {\ensuremath{\pi^{+}}\xspace}
\newcommand{\pim}          {\ensuremath{\pi^{-}}\xspace}
\newcommand{\kap}          {\ensuremath{\rm{K}^{+}}\xspace}
\newcommand{\kam}          {\ensuremath{\rm{K}^{-}}\xspace}
\newcommand{\pbar}         {\ensuremath{\rm\overline{p}}\xspace}
\newcommand{\kzero}        {\ensuremath{{\rm K}^{0}_{\rm{S}}}\xspace}
\newcommand{\lmb}          {\ensuremath{\Lambda}\xspace}
\newcommand{\almb}         {\ensuremath{\overline{\Lambda}}\xspace}
\newcommand{\Om}           {\ensuremath{\Omega^-}\xspace}
\newcommand{\Mo}           {\ensuremath{\overline{\Omega}^+}\xspace}
\newcommand{\X}            {\ensuremath{\Xi^-}\xspace}
\newcommand{\Ix}           {\ensuremath{\overline{\Xi}^+}\xspace}
\newcommand{\Xis}          {\ensuremath{\Xi^{\pm}}\xspace}
\newcommand{\Oms}          {\ensuremath{\Omega^{\pm}}\xspace}
\newcommand{\degree}       {\ensuremath{^{\rm o}}\xspace}

\begin{titlepage}
\PHyear{2021}       
\PHnumber{205}      
\PHdate{06 October}  

\title{Measurement of inclusive charged-particle b-jet production in \pp and \pPb collisions at \fivenn}
\ShortTitle{Production of b jets in \pp and \pPb collisions}   

\Collaboration{ALICE Collaboration\thanks{See Appendix~\ref{app:collab} for the list of collaboration members}}
\ShortAuthor{ALICE Collaboration} 

\begin{abstract}

A measurement of the inclusive b-jet production cross section is presented in \pp and \pPb collisions at \fivenn, using data collected with the ALICE detector at the LHC.
The jets were reconstructed in the central rapidity region $|\eta|<0.5$ from charged particles using the anti-\ensuremath{k_{\rm T}}\xspace algorithm with resolution parameter $R=0.4$. 
Identification of b jets exploits the long lifetime of b hadrons, using the properties of secondary vertices and impact parameter distributions. 
The \pT-differential inclusive production cross section of b jets, as well as the corresponding inclusive b-jet fraction, are reported for pp and \pPb collisions in the jet transverse momentum range $10 \le \pTchjet \le 100$\,\GeVc, together with the nuclear modification factor, \RpPbBjet. The analysis thus extends the lower \pT limit of b-jet measurements at the LHC.
The nuclear modification factor is found to be consistent with unity, indicating that the production of b jets in p--Pb at \fivenn is not affected by cold nuclear matter effects within the current precision. The measurements are well reproduced by POWHEG NLO pQCD calculations with PYTHIA fragmentation.

\end{abstract}
\end{titlepage}

\setcounter{page}{2} 

\section{Introduction}
\label{sec:intro}

Charm and beauty quarks arise from hard scattering processes with large four-momentum transfer ($Q^{2}$). In the subsequent hadronization process they lose their initial virtuality and produce short-lived heavy-flavor hadrons, which can be reconstructed either through their weak hadronic decays or indirectly via their semi-leptonic decay channels.
In case of proton--proton collisions, the inclusive production cross section of heavy-flavor hadrons can be calculated with perturbative quantum chromodynamics (QCD) using the factorization approach, which assumes that the collision process can be described by a convolution of parton distribution functions (PDFs), a short-distance parton-level cross section, and a fragmentation function.
This factorization was proven to be valid at the leading power of $Q$~\cite{ColSopSter:1988}, as well as the leading power corrections $O(1/Q^{2})$~\cite{QiuSter:1991a,QiuSter:1991b}. 
The concept of the QCD factorization is often extrapolated to proton--nucleus collisions by replacing the usual PDFs with nuclear PDFs (nPDFs), while keeping the short-distance parton-level cross section and the fragmentation function the same~\cite{EPS09:2009pA,Eskola:2008pA,Eskola:1999pA,DeFlorian:2004pA,Hirai:2007pA,Khalek:2020pA}. However, there are also additional phenomena which may or may not be incorporated into the nPDFs, for instance soft gluon interactions between the incoming and/or outgoing hadrons causing $k_{\rm T}$-broadening and energy loss of partons in the cold nuclear matter~\cite{Fujii:2013yja,Mangano:1991jk,Sharma:2009hn,Kang:2014hha}. These effects  may break the QCD factorization in nuclear collisions, thus making the nPDFs process dependent.
They are often accounted for as extra modification factors or convolutions with extra functions in various models~\cite{QiuVitev:2006pA, KopelRauf:2004pA,KoRaTarJo:2003pA,JoKoTar:2003pA}. The differences between the factorization of proton--proton and proton--nucleus collisions
are in general referred to as cold-nuclear-matter (CNM) effects. The overall impact of the CNM effects on the resulting \pT-differential inclusive production cross section spectrum can be quantified by means of the nuclear modification factor, defined as the ratio of the particle yield measured in proton--nucleus collisions and the expected yield that would be obtained from a superposition of independent \pp\ collisions. 
The sensitivity of heavy-flavor probes to CNM effects can be expected to differ from that of light-flavor probes due to the mass-dependent jet fragmentation~\cite{Dokshitzer:1991fd,Andronic:2015wma,Armesto:2003jh}.

Small collision systems such as pp or p--A provide a natural reference for the more complex nucleus--nucleus collisions.
Nuclear matter in these ultra-relativistic heavy-ion collisions can reach extremely high energy densities and temperatures, and transform into its hot and dense deconfined phase, the quark--gluon plasma (QGP)~\cite{Cabibbo1975,Chapline:1976gy,Shuryak:1978ij}. Initial parton showers interact with the medium via collisional and radiative processes that cause  dissipation and redistribution of energy inside the parton shower. 
This results in the suppression of high-\pT hadrons and jets~\cite{Gyulassy:1999zd,Wang:2002ri,Adcox:2001jp,Aamodt:2010jd,CMS:2012aa} in
nucleus--nucleus collisions and the modification of the jet
substructure~\cite{Sirunyan:2017bsd,Acharya:2019djg,Milhano:2017nzm,Mehtar-Tani:2016aco}, the so-called jet quenching. 
Since heavy-flavor quarks are mainly produced in initial hard processes and since their numbers remain largely unchanged in the later stages of the reaction~\cite{Braun-Munzinger:2007fth,STAR:2018zdy}, they provide a unique opportunity to study the space--time evolution of the QGP. In this context, small collision systems represent an important test for theoretical models that account for the system-size-dependent evolution of the QGP signatures as well as CNM effects. Understanding CNM effects is therefore essential for the accurate quantification of the effects of a hot and dense medium in heavy-ion measurements.

The reconstruction of jets containing heavy-flavor hadrons provides more direct access to the primary heavy-flavor parton kinematics than an inclusive measurement of heavy-flavor hadrons. By measuring heavy-flavor jets, production and fragmentation effects can be studied separately. The ALICE Collaboration reported production of charm-tagged jets in \pp collisions at $\sqrt{s}=7$\,TeV~\cite{Acharya:2019zup}.
Measurements of beauty-tagged jets (b jets) in \pp and \pPb collisions were performed by the CMS experiment~\cite{Khachatryan:2015sva}. They reported the nuclear modification factor for b jets with transverse momentum larger than 50\,\GeVc.
The ALICE detector has excellent tracking capabilities for low-\pT charged particles, which makes it possible to measure b jets at low transverse momenta. 
This provides a unique opportunity at the LHC to study nuclear modification of b jets down to the region where the energy scale of the jets is of similar magnitude compared to the b-quark mass, which increases sensitivity to mass dependent effects.
 In this paper, we present the first measurement of inclusive charged-particle b-jet \pT-differential cross section and the b-jet fraction, down to jet-transverse momentum $\pTchjet=10$\,\GeVc in \pPb and \pp collisions at \fivenn. 
The measured \pT distributions were used to obtain the nuclear modification factor of b jets, \RpPbBjet, in the transverse momentum range $10 \le \pTchjet \le 100$\,\GeVc.

The paper is organized as follows: the next section introduces the experimental setup and data sets used for these measurements. 
Jet reconstruction and the b-jet tagging procedures are described in Sec.~\ref{sec:btag}. Section~\ref{sec:Corrections} deals with the correction steps that were applied in the analysis. These include corrections for b-jet tagging efficiency, b-jet tagging  purity,
 and unfolding  of the jet momentum smearing due to underlying event fluctuations and instrumental effects. Systematic uncertainties are discussed in Sec.~\ref{sec:Systematic}. Section~\ref{sec:Results} is devoted to the discussion of the final results. The paper is summarized in Sec.~\ref{sec:summary}.

\section{Experimental setup and data sets}\label{sec:DATA}

 The ALICE detector~\cite{Aamodt:2008zz,Alicedet:ijmpa} consists of a central barrel, a forward muon arm, and a set of forward detectors that are used for triggering and event characterization. The central barrel hosts detection systems that provide tracking and particle identification. The most important ones for this analysis are the Inner Tracking System (\ITS) and the Time Projection Chamber (\TPC). The \ITS is a 6-layer silicon tracker, which allows for precise reconstruction of primary interaction and secondary decay vertices. 
 The two innermost layers of the \ITS are formed by the Silicon Pixel Detector (SPD).
 All detectors of the central barrel are placed in a solenoidal magnet that provides a field of 0.5\,T along the beam direction.

The present analysis is based on the \pPb and \pp collisions at \fivenn taken by ALICE in 2016 and 2017, respectively.  For \pPb collisions, the beam energies of colliding protons and Pb nuclei were asymmetric: the protons had 4 TeV, while Pb nuclei had an energy of 1.59\,TeV per nucleon. This resulted in the laboratory frame in a rapidity ($y$) shift of the nucleon--nucleon center-of-mass system by $\Delta y=0.465$  in the direction of the proton beam.

The main triggering device for the data sets used here is the \VZERO detector~\cite{Abbas:2013taa}, consisting of two scintillator arrays \VZEROA and \VZEROC. They cover the full azimuth angle in the forward and backward pseudorapidity ranges $2.8<\eta<5.1$ and  $-3.7<\eta<-1.7$, respectively. The minimum bias trigger (MB) is defined by a coincidence of \VZEROA and \VZEROC signals.
Timing of the \VZEROA and \VZEROC signals is also used to reject background from beam--gas interactions.

Pile-up events constitute less than 1\% (0.5\%) of triggered events in pp (\pPb) collisions. They were identified and rejected using an algorithm that utilizes track segments, formed by hits in the SPD, to recognize events with multiple primary vertices.  The remaining undetected pile-up events constitute a negligible fraction of the analysed sample.

The \pPb data set corresponds to an integrated luminosity of $\mathcal{L}_{\rm pPb} = (298\pm11)\,{\upmu}$b$^{-1}$ ($624\times10^{6}$ MB events)~\cite{Abelev:2014epa}, and the \pp data set to $\mathcal{L}_{\rm pp} =(18.9\pm0.4)$\,nb$^{-1}$ ($968\times10^{6}$ MB events)~\cite{ALICE-PUBLIC-2016-005}. Only events with the location of the reconstructed primary vertex along the beam axis within $|z_{\rm vtx}| < 10$\,cm were retained to assure a uniform detector coverage at midrapidity.

\section{Jet reconstruction and b-jet identification}
\label{sec:btag}

The analysis uses high-quality tracks~\cite{Acharya:2019zup} reconstructed in the pseudorapidity range $|\eta_{\rm track}|<0.9$ that have at least one hit in either of the two \SPD layers.
In the regions where the \SPD was inefficient, high-quality tracks were supplemented with  complementary  tracks  that  do  not  have  a  hit  in  the  \SPD, to achieve azimuthal uniformity in the tracking acceptance. The momentum resolution of complementary tracks is improved by constraining the origin of the track to the primary vertex.
Complementary tracks constitute about 3.5\% of  all  primary tracks. The tracking efficiency for primary tracks with $\pT>1 $\,\GeVc varies with \pT between 70 and 85\%. Primary-track  momentum  resolution  is about  0.7\%  at $\pT=1$\,\GeVc,  1.6\%  at $\pT=10$\,\GeVc,  and  4\%  at $\pT=50$\,\GeVc. The spatial resolution of the track impact parameter with respect to the primary vertex is better than $75\,\upmu$m for charged-particle tracks with transverse momentum $\pT > 1$\,\GeVc\ and better than $20\,\upmu$m for tracks with $\pT > 20$\,\GeVc\ ~\cite{Acharya:2019zup,Alicedet:ijmpa}. More information about the track selection can be found in Ref.~\cite{Acharya:2019zup}.

Jets were reconstructed using the infrared and collinear safe anti-$k_{\rm T}$ algorithm~\cite{Cacciari:2008gp} from the FastJet package~\cite{Cacciari:2011ma}. The resolution parameter was set to $R=0.4$, which ensures that most of the momentum of the initial parton (approximately 70\% to 90\% in the range of the current measurement) falls within the jet cone~\cite{Chatrchyan:2012mec}. The jets were constructed from charged particles having $p_{\rm T,track}>0.15$\,\GeVc and pseudorapidity $|\eta_{\rm track}|<0.9$. Their four-momenta were combined using the \pT recombination scheme, which considers all particles to be massless~\cite{Cacciari:2011ma}.
The pseudorapidity coverage of the reconstructed jets was constrained to $|\eta _{\rm jet}|<0.9-R=0.5$ to select only jets that are fully contained within the TPC acceptance.


The reconstructed transverse momentum for jets $\pTchjet^{\rm reco}$ is obtained using the measured transverse momentum of charged-particle jets $\pTchjet^{\rm raw}$, corrected for the mean contribution of the underlying event using the formula
$\pTchjet^{\rm reco}=\pTchjet^{\rm raw} \, - \, \rho \times A_{\rm jet}$~\cite{Cacciari:2008rho}.
Here $A_{\rm jet}$ denotes the area of the jet and $\rho$ is the mean underlying event \pT density. The mean underlying event \pT density was calculated on an event-by-event basis using the estimator introduced by CMS~\cite{Chatrchyan:2012tt}.

Identification of b jets is based on kinematic variables related to the lifetime of b-hadrons ($c\tau \approx 500$\,$\upmu$m), and the large impact parameter of beauty-hadron-decay daughters. Several discriminator variables were defined and applied in two distinct b-jet tagging methods that are presented in this paper, the impact parameter (IP) method (based on the distance of closest approach, DCA, of the individual jet tracks to the primary vertex), and the displaced secondary vertex (SV) method (based on the topology of a reconstructed secondary vertex using a subset of the jet tracks). The tagging of the b-jet candidates utilizes global tracks only and exploits the high spatial resolution of the SPD.
The b-jet \pT-differential spectra were separately obtained with the two tagging algorithms and eventually combined to improve the accuracy of the measurement. While the IP method generally provides better b-jet tagging efficiency, the SV method has been proven to be more stable at low \pT.
Both methods are discussed in detail below. For more information on b-jet tagging algorithms, the reader may refer to Refs.~\cite{Chatrchyan:2012jua,MontejoBerlingen2016,Abazov:2010ab}.

\subsection{b-jet tagging based on impact parameter}\label{sec:TrackCounting}

The impact parameter of a track can be measured either in three dimensions or in the projection on the plane perpendicular to the beam axis. This analysis used the latter definition (denoted $d_{xy}$) to exploit the better resolution of the ITS in this plane.

\begin{figure}[h!]%
\centering
{\includegraphics[width=0.7\columnwidth]{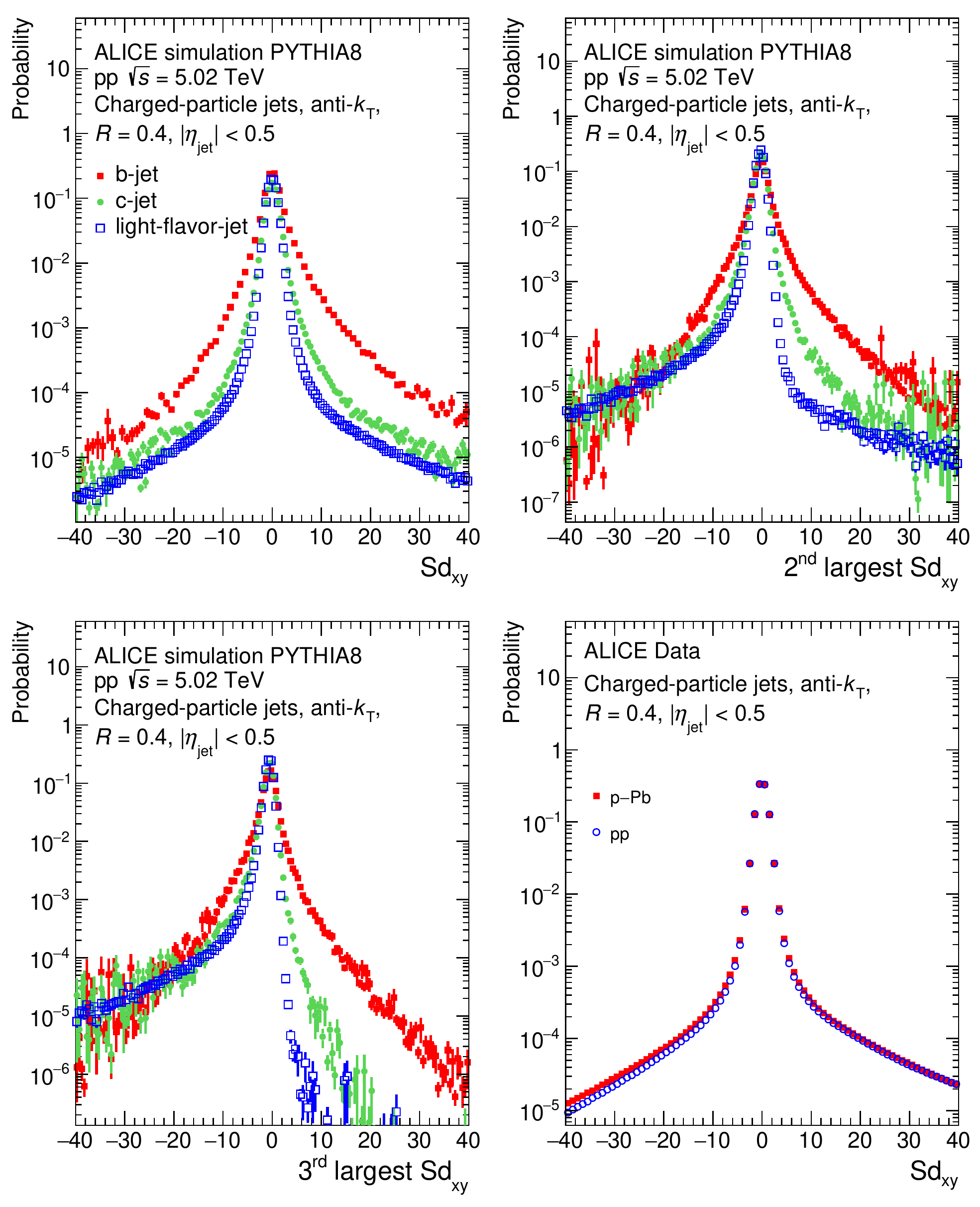}}%
\caption{Top left: The normalized impact parameter significance ($Sd_{xy}$) distribution for all tracks inside light-flavor, charm, and beauty jets as determined from PYTHIA~8 (Monash 2013 tune~\cite{Skands:2014pea}) detector-level simulations. Top right: The distribution of the second largest impact parameter significance in the jet. Bottom left: The distribution of the third largest $Sd_{xy}$ in the jet. Bottom right: Distribution of $Sd_{xy}$ for data in \pp and \pPb collisions.}
\label{fig:IPSflavor}
\end{figure}

The sign of the impact parameter is determined as the sign of the scalar product of the jet axis and the impact parameter vector pointing from the primary vertex to the point of closest approach.
Tracks originating from a secondary vertex tend to have positive impact parameter values because of the mother particle decay length.
On the other hand, the tracks originating from the primary vertex can have both positive and negative impact parameter values due to finite resolution which smears the impact parameter of primary tracks symmetrically around the primary vertex.
Discrimination among different jet flavors was based on the impact parameter significance ($Sd_{xy}$), defined as the ratio of the impact parameter to its estimated resolution.
The impact parameter resolution largely depends on the $\eta$ and \pt of the tracks. Figure~\ref{fig:IPSflavor} (top left) shows the probability distribution of the impact parameter significance for tracks belonging to different jet flavors, as determined 
from a detector-level simulation, where PYTHIA~8 Monash 2013~\cite{Sjostrand:2006za} events were processed with an ALICE GEANT~3-based particle transport model~\cite{Brun:1994aa}.

On average, tracks associated to b jets have larger $Sd_{xy}$ values when
compared to c jets and light-flavor jets. This means that the impact parameter has a strong discriminating power in distinguishing between the different jet flavors.

This analysis uses the track counting algorithm~\cite{Chatrchyan:2012jua}, which arranges the $Sd_{xy}$ values of tracks in a jet in descending order. A jet was tagged as a b jet if the second largest impact parameter significance value (see Fig.~\ref{fig:IPSflavor} top right)  was greater than a certain threshold parameter $Sd_{xy}^{\rm min}$. The default threshold parameter that was chosen in this analysis is $Sd_{xy}^{\rm min}=2.5$, which gives an average tagging efficiency of $55\%$ with average purity of $42\%$ for b jets with $20<\pTchjet^{\rm reco}<40$\,\GeVc. This choice provided an optimum balance between good efficiency and good background rejection. Discrimination based on the tracks with the first largest as well as the third largest impact parameter significance value (see Fig.~\ref{fig:IPSflavor} bottom left)  were used for consistency checks.

The purity and b-jet tagging efficiency of the selected b-jet sample presented in Sec.~\ref{sec:TaggingEfficiencyIP} were determined using the jet probability algorithm~\cite{Abazov:2010ab,Abdallah:2002xm,Chatrchyan:2012jua}.
This algorithm evaluates a combined impact parameter significance of tracks inside the jet and estimates a likelihood that all tracks associated with the jet originated from the primary vertex.

\begin{figure}[h!]%
\centering
{\includegraphics[width=.5\columnwidth]{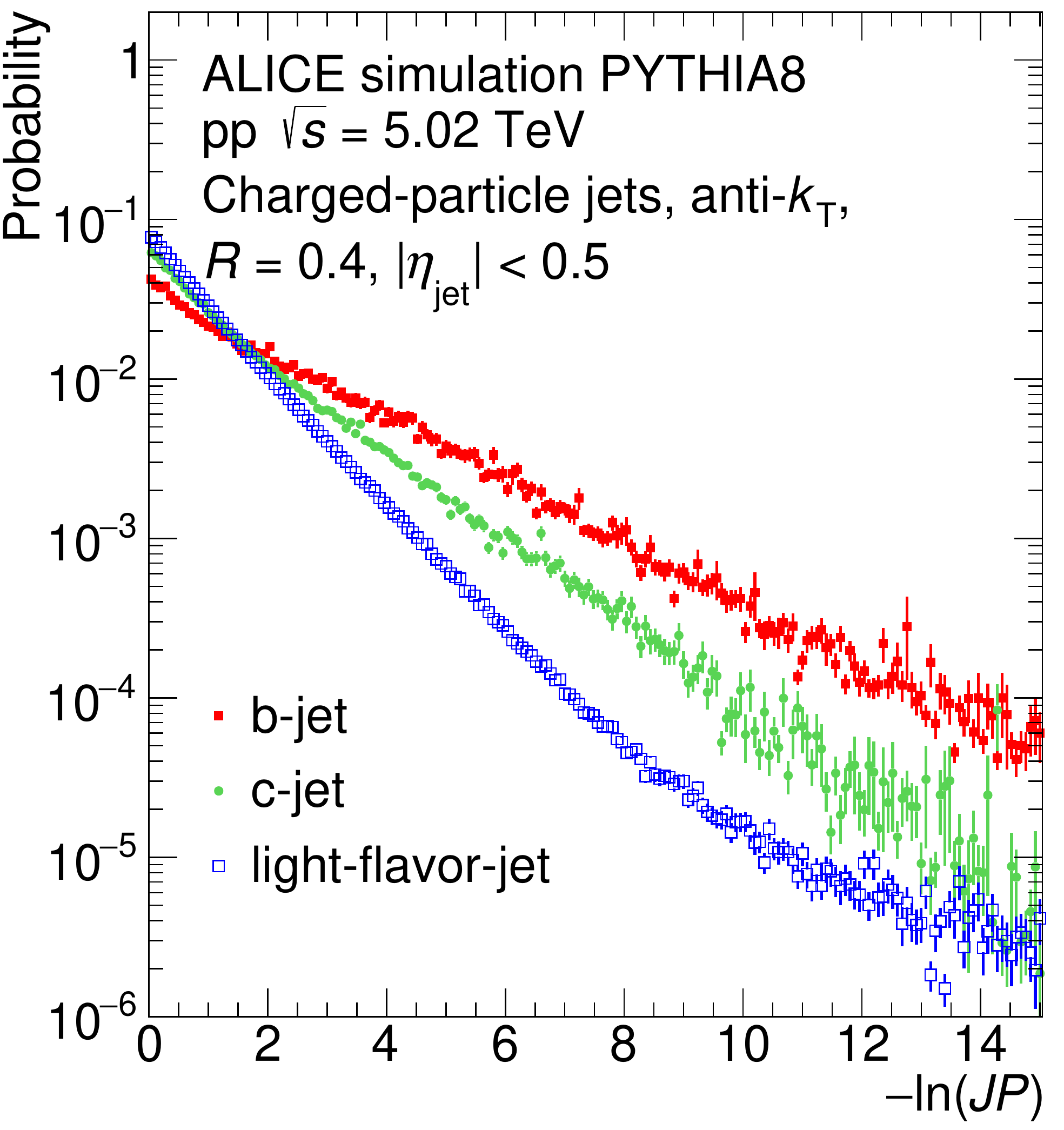}}%
\caption{The logarithmic jet probability $-\ln(JP)$ distribution for light-flavor, charm, and beauty jets in \pp collisions at \five.}
\label{fig:LnJP}
\end{figure}

Reconstructed tracks were classified based on different geometric and tracking features.
The algorithm defines a resolution function $R_{\rm IP}$ for each category, by fitting the negative side of the signed $Sd_{xy}$ distribution (bottom right panel of Fig.~\ref{fig:IPSflavor}). This fit is carried out on the negative part of the distribution because in this range it is predominantly populated by primary tracks originating from the primary vertex.

The resolution functions corresponding to the different track categories were used to calculate the track probability $P_{\rm tr}$. This $P_{\rm tr}$ corresponds to the probability that a high-quality jet constituent track with an impact parameter significance $Sd_{xy}$ is coming from the primary vertex:
\begin{equation}
 P_{\rm tr}(Sd_{xy}) = \frac{\int _{-\infty}^{-|Sd_{xy}|} R_{\rm IP}(S){\rm d}S}{\int _{-\infty}^{0} R_{\rm IP}(S){\rm d}S}\,,
 \label{eq:TrackProb}
\end{equation}
where the integration is done over the negative side of the impact parameter significance distribution. A large impact parameter value results in a small $P_{\rm tr}$. 

The jet probability ($JP$) is then calculated by combining the $P_{\rm tr}$ values of tracks within a given jet according to the equation~\cite{Abazov:2010ab,Abdallah:2002xm,Chatrchyan:2012jua}:
\begin{equation}
JP = \prod \times  \sum_{k=0}^{N_{\rm track}-1} \frac{(-\log \prod)^{k}}{k!} \;\;\; \mbox{,\,where} \;\;\; \prod = \prod_{i=1}^{N_{\rm track}} P_{{\rm tr}, i}\,.
\label{eq:JetProb}
\end{equation}
Only tracks with positive $Sd_{xy}$ are selected to calculate the jet probability. 
The $JP$ discriminates over different jet flavors only in a very narrow interval ($0<JP<0.2$). This distribution is therefore not convenient for discrimination. For this reason, the $-\ln (JP)$ quantity was used as a discriminator in our analysis to determine the b-jet tagging efficiency using a data-driven method. As shown in 
Fig.~\ref{fig:LnJP}, the $-\ln (JP)$ decreases much faster for light-flavor jets and c jets when compared to b jets, allowing for an effective statistical discrimination of b jets. 

\subsection{b-jet tagging based on secondary vertex reconstruction}\label{sec:SecondaryVertex}

Secondary vertices (SV), where the weak decay of the beauty hadrons take place, are in most cases well displaced from the primary vertex of the collision due to the lifetime of beauty hadrons. 
Beauty hadrons primarily decay to non-prompt charm particles which typically have
similarly long lifetime.
The SV algorithm reconstructs the secondary vertices inside the jets from triplets of jet-constituent tracks.
This choice was motivated by the typical decay patterns of beauty hadrons.
From all of these reconstructed secondary vertices, this algorithm selects for the b-jet tagging the vertex that is most displaced.
The vertex reconstruction quality is described by the dispersion of the reconstructed secondary vertex, $\sigma _{\rm SV} = \sqrt{d_1^2 + d_2^2 + d_3^2}$, where $d_{1,2,3}$ are the distances of closest approach of the three tracks to the secondary vertex.

This algorithm uses the decay length $L_{xy}$ as a discriminator. The decay length is the distance between the primary vertex and the secondary vertex measured in the plane transverse to the beam axis. The significance is then defined by dividing $L_{xy}$ by its uncertainty, $SL_{xy}=L_{xy}/\sigma _{L_{xy}}$.
The b-tagging is then performed by considering both the SV dispersion $\sigma _{\rm SV}$ and the decay length significance $SL_{xy}$. 

\begin{figure}[h!]%
\centering
 \includegraphics[width=.5\linewidth]{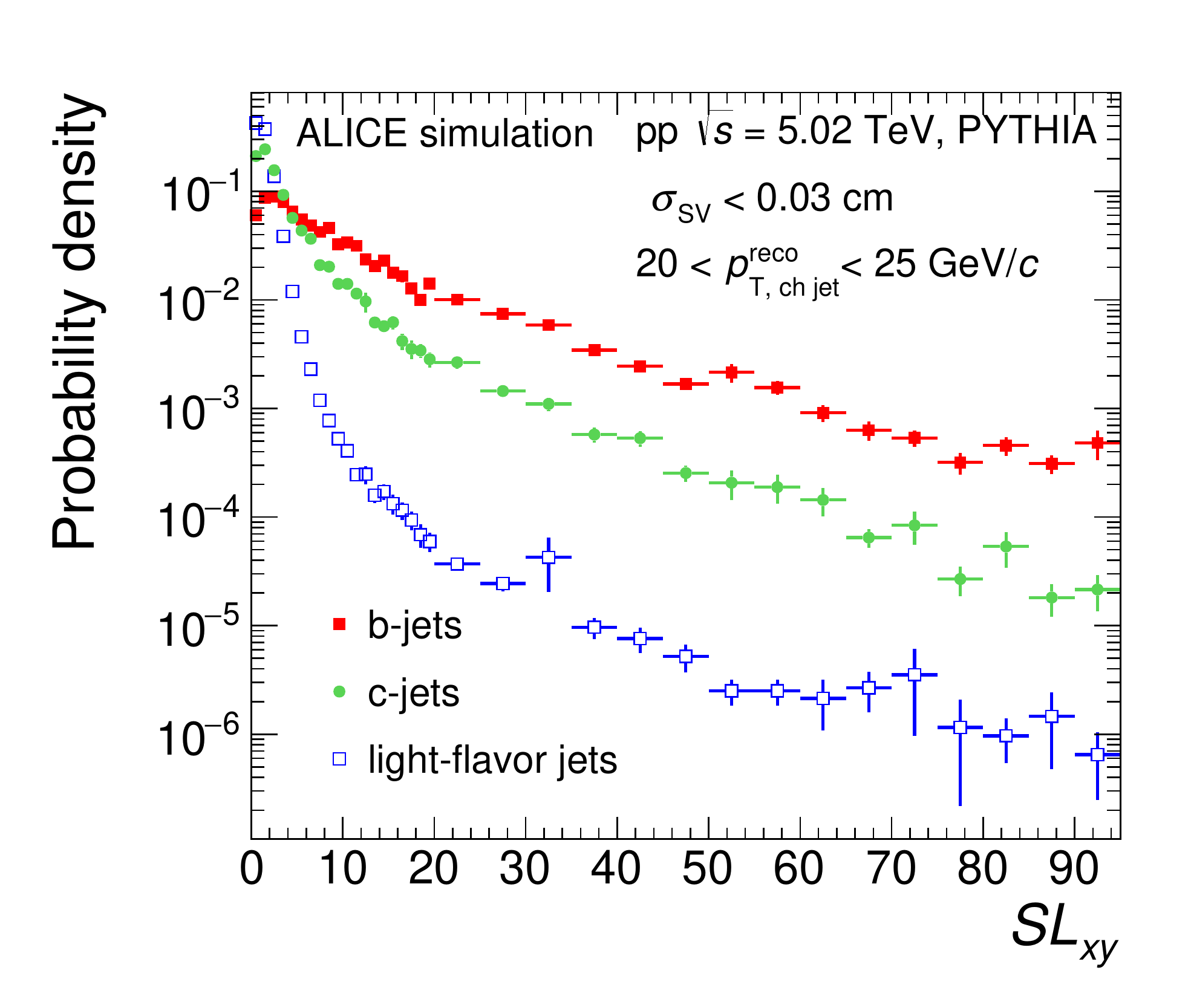}%
 \includegraphics[width=.5\linewidth]{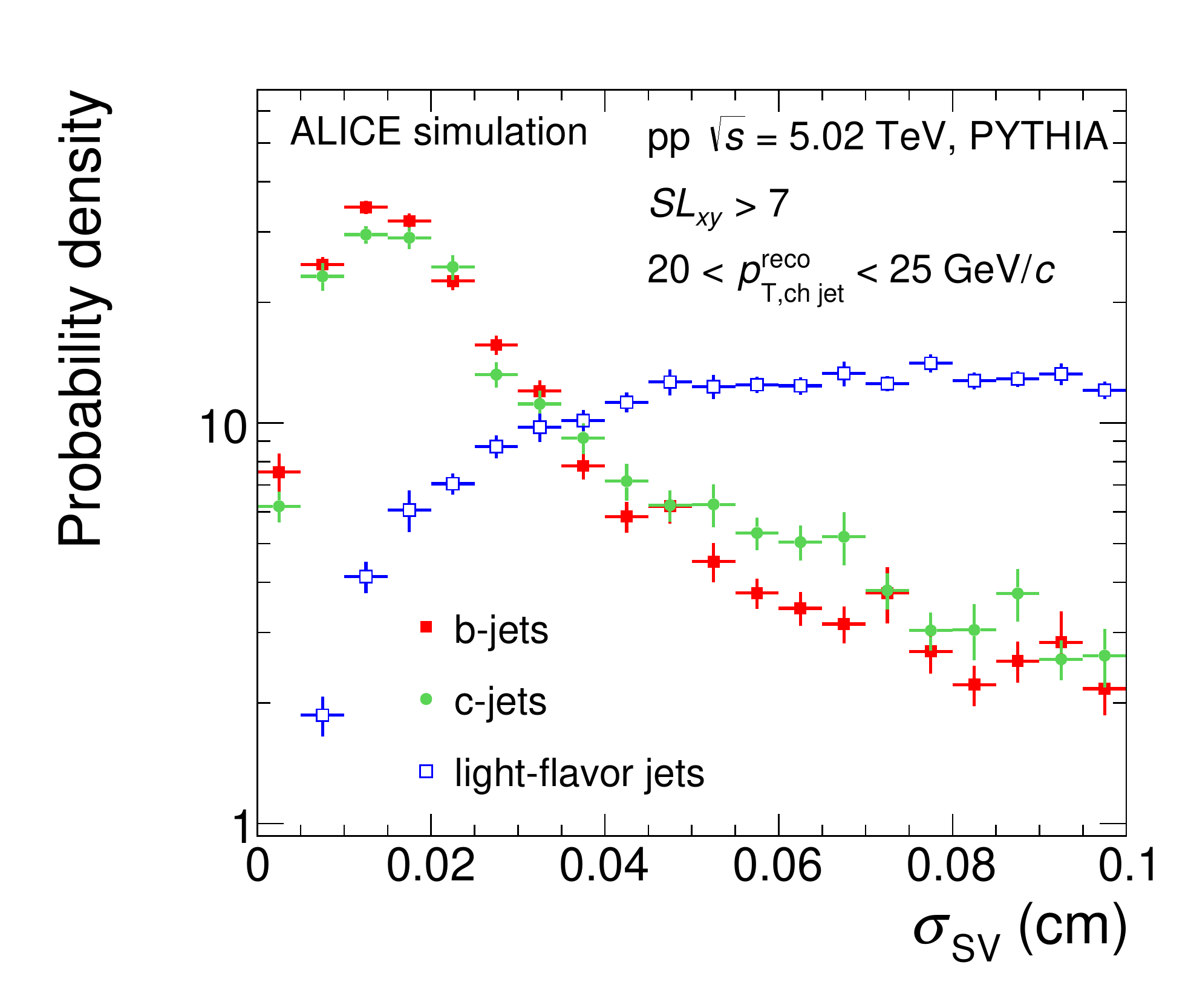}
\caption{Distributions of the tagging discriminators used in the SV method, $SL_{xy}$ (left) and $\sigma_{\rm SV}$ (right), for b jets, c jets, and light-flavor jets as obtained from a MC simulation of the ALICE apparatus, using PYTHIA as an event generator.}
\label{fig:SLxySigmaSV_PYTHIA}
\end{figure}

\begin{figure}[h!]%
\centering
 \includegraphics[width=.5\linewidth]{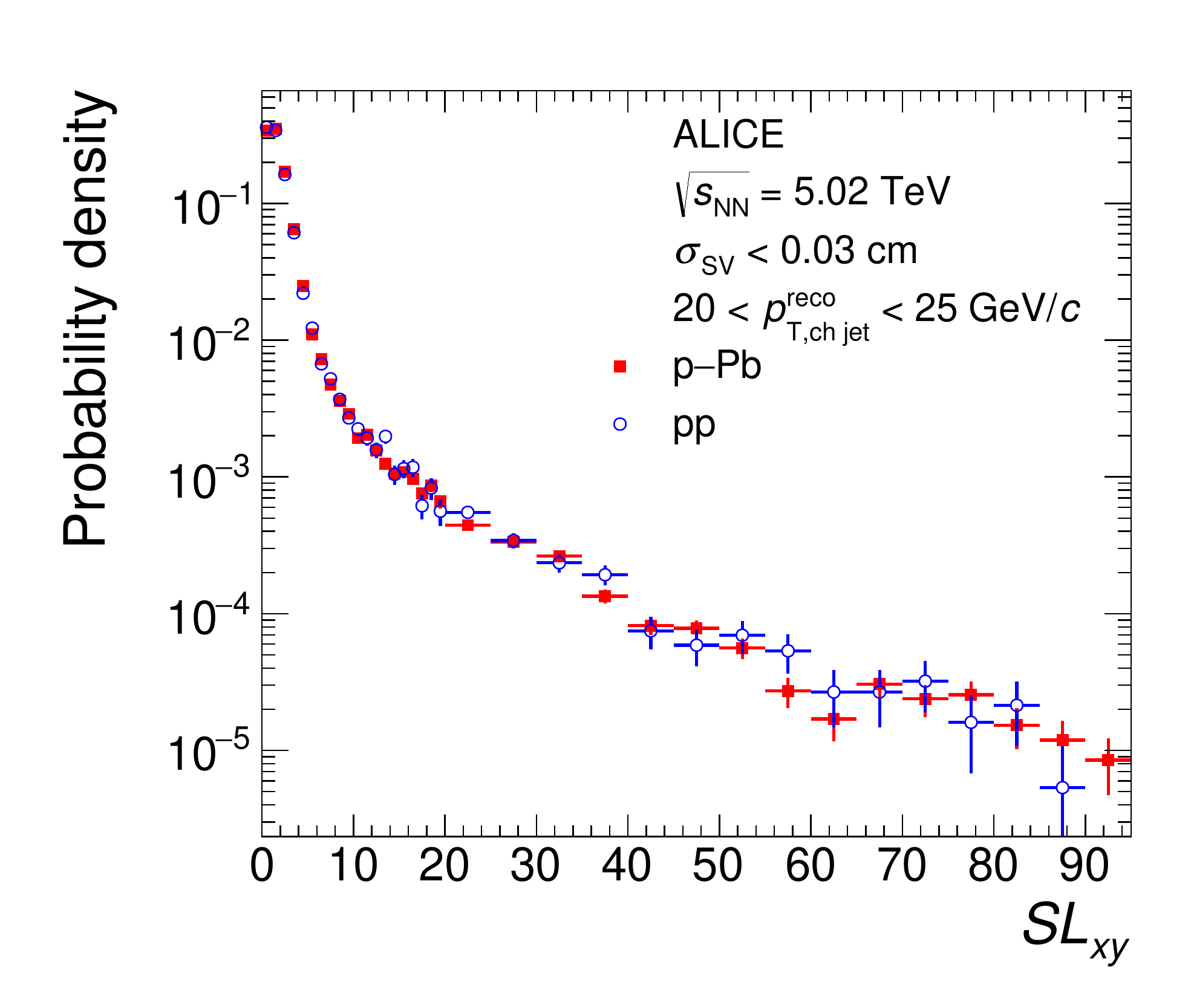}%
 \includegraphics[width=.5\linewidth]{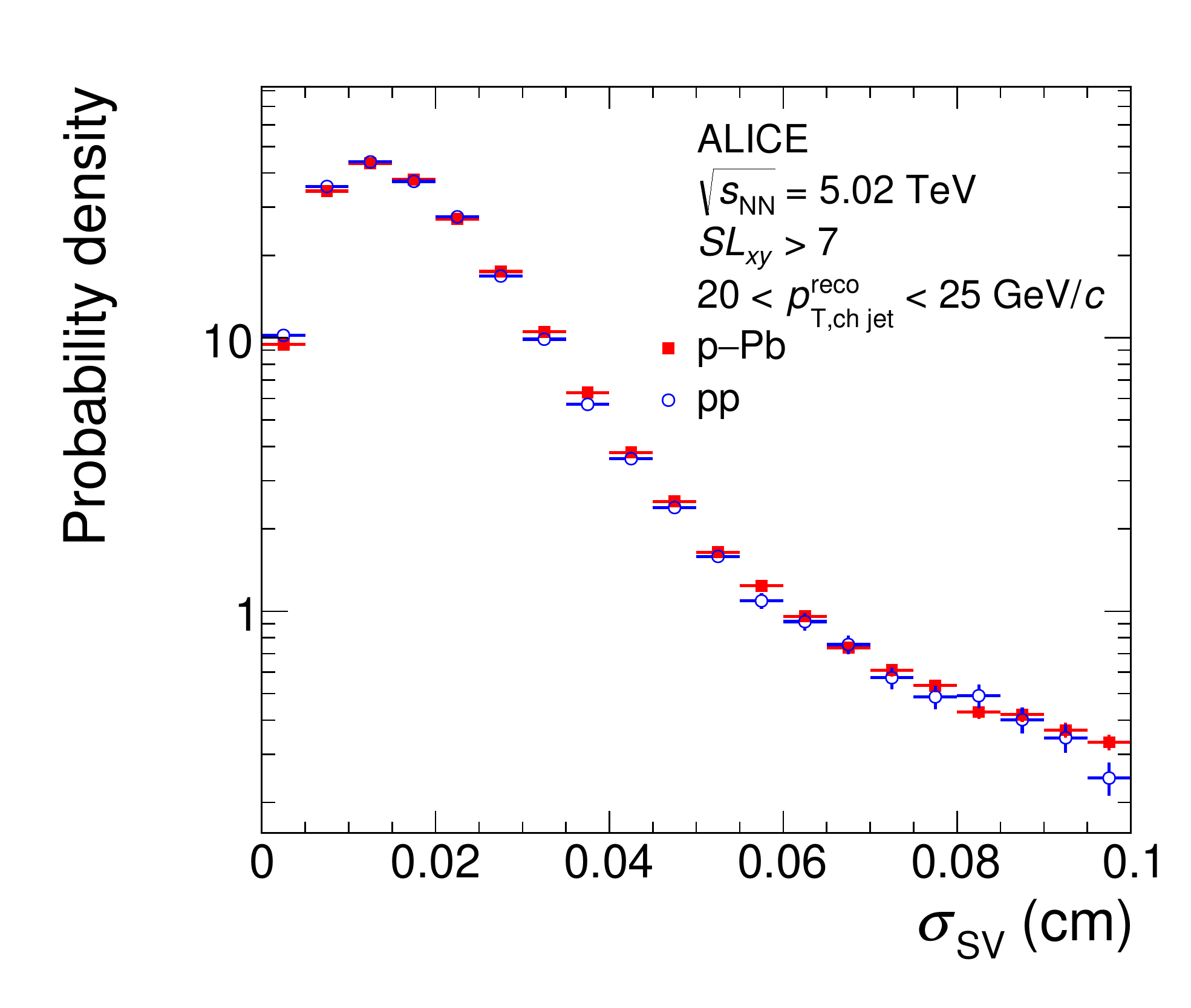}
\caption{Distributions of the tagging discriminators used in the SV method, $SL_{xy}$ (left) and $\sigma_{\rm SV}$ (right), for \pp and \pPb collisions.}
\label{fig:SLxySigmaSV}
\end{figure}

The default operating point of the tagging in the analysis is $SL_{xy}>7$ and $\sigma _{\rm SV}<0.03$\,cm.  These selection values were determined by  optimizing for high b-tagging efficiency and low c-quark and light-flavor mistagging rates based on simulations. Figure~\ref{fig:SLxySigmaSV_PYTHIA} shows examples of the $SL_{xy}$ and $\sigma _{\rm SV}$ distributions for jets having different flavors as obtained from PYTHIA 8 simulations using the Monash tune~\cite{Skands:2014pea} followed by an ALICE detector level MC simulation and reconstruction.
Figure~\ref{fig:SLxySigmaSV} shows examples of the $SL_{xy}$ and $\sigma _{\rm SV}$ probability distributions in \pp\ and \pPb\ collision data.

\section{Corrections to the b-tagged jet spectrum}\label{sec:Corrections}

The raw \pT spectrum of b-jet candidates ($\mathrm{d} N^{\rm tagged}/\mathrm{d}\pTchjet^{\rm reco}$) that was obtained after applying the tagging algorithms was corrected for the b-jet tagging efficiency, $\epsilon_{\rm b}$, and the purity of the selected b-jet sample, $P_{\rm b}$,
\begin{equation}
    \frac{{\rm d} N^{\rm b\,\,jet}_{\rm det.\,level}}{{\rm d} \pTchjet^{\rm reco}} = \frac{\mathrm{d} N^{\rm tagged}}{ \mathrm{d} \pTchjet^{\rm reco}} \times \frac{P_{\rm b}}{\epsilon _{\rm b}} \,. 
\label{eq:rawspeccorr}
\end{equation} 
The resulting spectrum is then corrected for jet reconstruction efficiency and momentum smearing due to detector effects and background fluctuations by means of unfolding.
All corrections are discussed below in detail.

\subsection{Tagging efficiency}\label{sec:TaggingEfficiency}
The b-jet tagging algorithms discussed in Section~\ref{sec:btag} do not identify all produced b jets. The probability that a given tagging algorithm correctly identifies a jet originating from a b quark as a b jet is called the tagging efficiency. Similarly, one can also define the mistagging efficiency as the probability that a jet originating from a charm quark or a light-flavor parton is falsely tagged as a b jet. 
The efficiency of a given algorithm for tagging or mistagging is defined as
\begin{equation}
\epsilon_i (\pTchjet^{\rm reco}) = \frac{N_{i}^{\rm tagged} (\pTchjet^{\rm reco})}{N_{i}^{\rm total} (\pTchjet^{\rm reco})}\ , 
\label{eq:TaggingEff}
\end{equation}
where $i$ is the jet flavor (b, c or light-flavor), $N_{i}^{\rm tagged}$ is the number of tagged $i$ jets, and $N_{i}^{\rm total}$ is the total number of $i$ jets.

\subsubsection{Tagging efficiency of the IP algorithm}
\label{sec:TaggingEfficiencyIP}

The tagging efficiency of the IP algorithm was estimated based on the semi-data-driven method outlined in Refs.~\cite{Khachatryan:2015sva,Chatrchyan:2012jua}, where the $-\ln(JP)$ distributions are fitted with a set of detector-level MC templates, which describe the shape of the jet probability distributions corresponding to b jets, c jets, and light-flavor jets. The templates for \pPb were obtained from a MC simulation based on the EPOS event generator~\cite{Pierog:2013ria} with embedded PYTHIA~6 events, where particles are propagated through a model of the ALICE detector using GEANT~3~\cite{Brun:1994aa}. The simulated events were then reconstructed as events in data.
The templates for \pp collisions were obtained similarly, using the PYTHIA~8 MC event generator.

\begin{figure}[h!]%
\centering

\subcaptionbox{\centering Untagged jets}{\includegraphics[width=.5\linewidth]{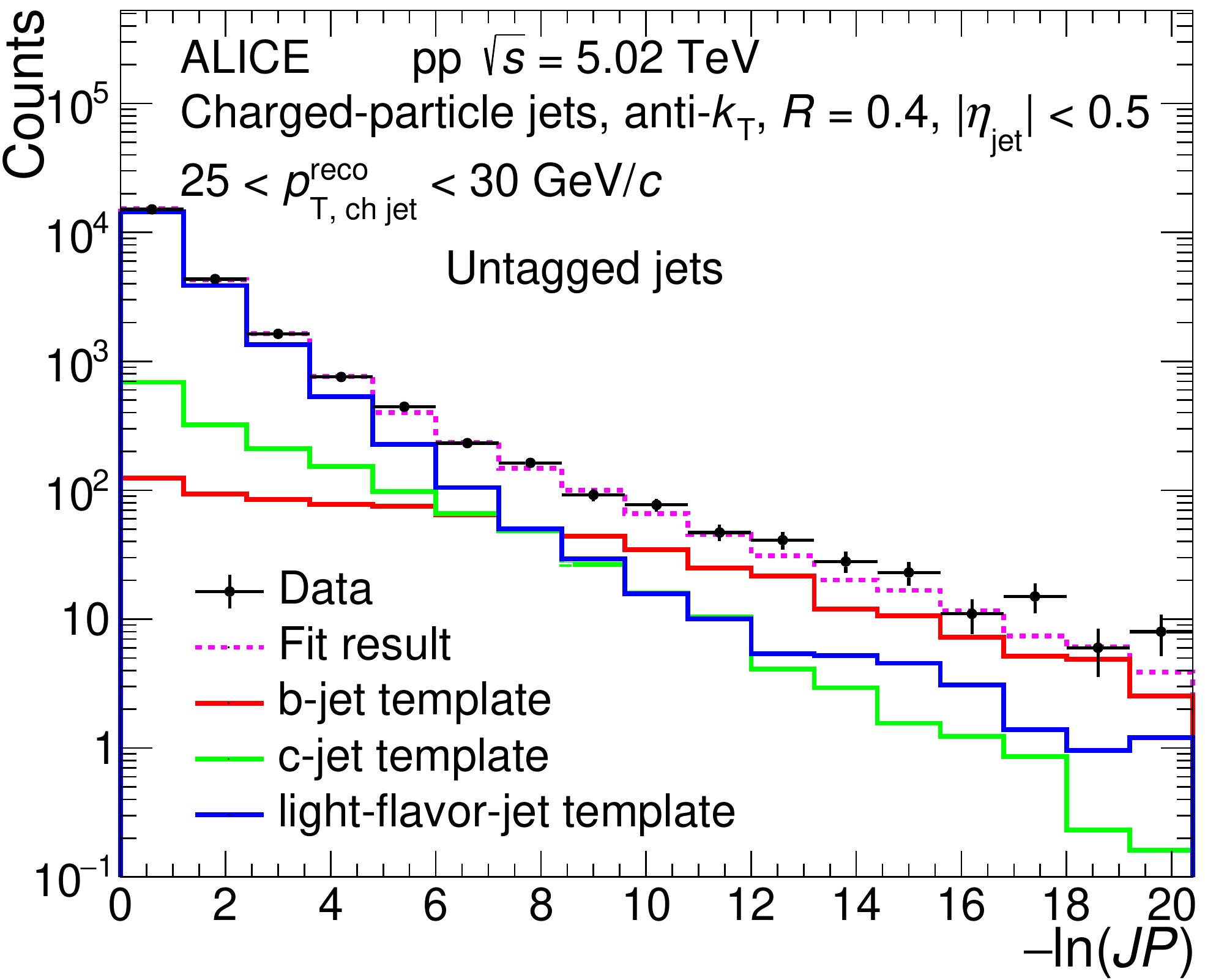}}%
\subcaptionbox{\centering Tagged jets}{\includegraphics[width=.5\linewidth]{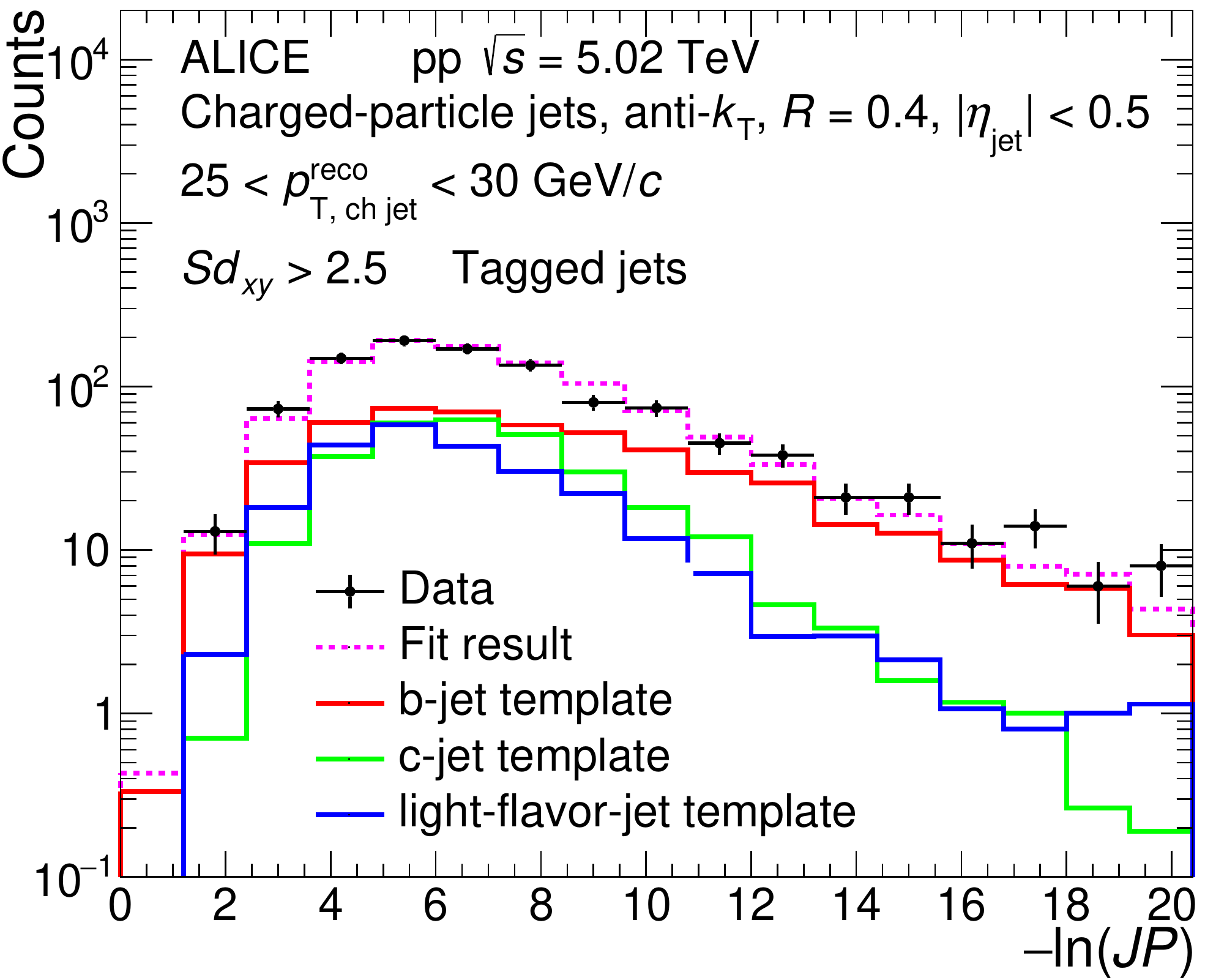}}%

\caption{Fit of the measured $-\ln(JP)$ discriminator distribution with a linear combination of b, c, and light-flavor jet templates for the untagged sample (left) and for the tagged sample (right).}
\label{fig:LTRT}
\end{figure}

Two jet samples were created: a sample that contains the jets satisfying the tagging requirement (tagged sample), and another sample that contains the inclusive jets before applying the tagging algorithm (untagged sample). The associated $-\ln(JP)$ distributions from data were fitted with the corresponding b, c, and light-flavor jet templates using a binned maximum likelihood fit. The fitting procedure was done separately for the tagged jet (with $Sd_{xy}^{\rm min} = 2.5$) and the inclusive (untagged) jet samples, see Fig.~\ref{fig:LTRT}.

The b-jet tagging efficiency is then obtained as the ratio of the number of identified b jets to the number of b jets before identification:
\begin{equation}
\epsilon_{\rm b} = \frac{C_{\rm b} \times f_{\rm b}^{\rm tag} \times N_{\rm data}^{\rm tag} }{f_{\rm b}^{\rm untag} \times N_{\rm data}^{\rm untag}}\,.
\end{equation}
Here $f_{\rm b}^{\rm untag}$ and $f_{\rm b}^{\rm tag}$ denote the b-jet fractions before and after tagging, respectively, which are extracted from the fits;  $N_{\rm data}^{\rm untag}$ and $N_{\rm data}^{\rm tag}$ give the numbers of jets before and after tagging, which were extracted from data; finally, $C_{\rm b}$ is the fraction of b jets for which the jet probability can be defined, i.e. b jets having at least two constituent tracks with positive $Sd_{xy}$. This factor was estimated from MC. The $C_{\rm b}$ is $\approx$80\% at 10\,\GeVc and increases to 98\% at 40\,\GeVc and remains at that value for $\pT > 40$\,\GeVc.

\begin{figure}[h!]%
\centering

{\includegraphics[width=.65\linewidth]{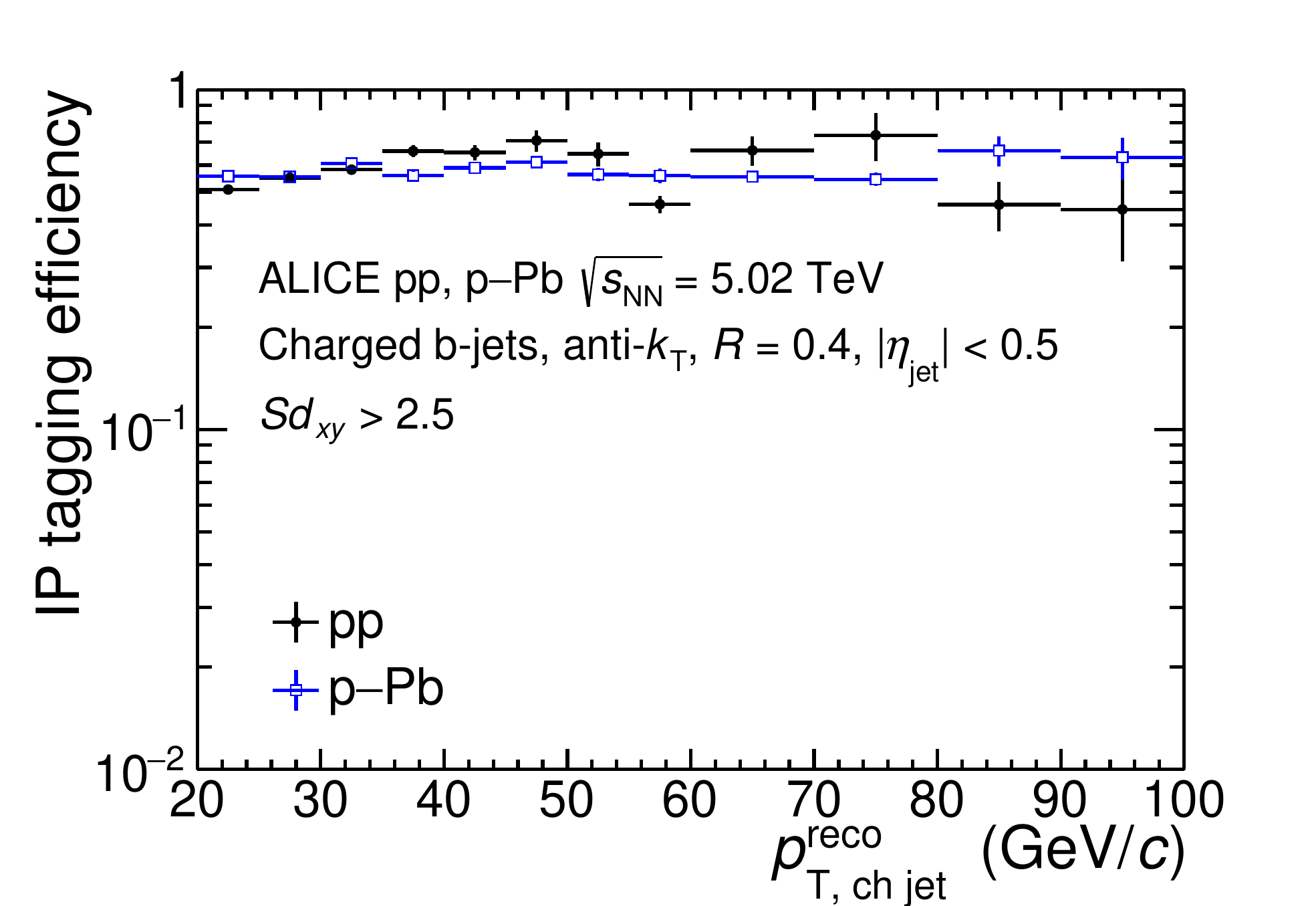}}%

\caption{The b-jet tagging efficiency extracted from the data-driven method using the IP algorithm in \pp and \pPb collisions.}
\label{fig:Efficiency_DATA_MC}
\end{figure}

Figure~\ref{fig:Efficiency_DATA_MC} shows the b-jet tagging efficiency of the IP method in \pp and \pPb collisions.
As an alternative for $-\ln(JP)$ in the template fitting, other discriminators were also used to check consistency and estimate systematic uncertainties.
The alternative discriminators were the jet mass distribution~\cite{Acharya:2017goa} and the distribution of energy fraction $f_E$ carried by the secondary vertex in the jet. Both of them provide results that are consistent with the standard analysis within one standard deviation. The systematic uncertainty on the tagging efficiency is estimated by fitting the $f_E$ distribution instead of $-\ln(JP)$. While $JP$ may be correlated with the IP, there is no such correlation in the case of $f_E$. The good match between efficiencies and purities obtained with the different methods excludes the possibility that any such correlation affects the results.
Finally, it is worth noting that the template fit procedure yields results with large systematic uncertainties for $\pTchjet < 20$\,\GeVc, so the interval between $10<\pTchjet<20$\,\GeVc was omitted in the IP analysis. 
The reason for these uncertainties is that the individual templates have rather similar shapes, causing instabilities in the fitting algorithm and thus reducing the discrimination power of the fit.

\subsubsection{Tagging efficiency of the SV algorithm}
\label{sec:TaggingEfficiencySV}

\begin{figure}[h!]
\includegraphics[width=.65\linewidth]{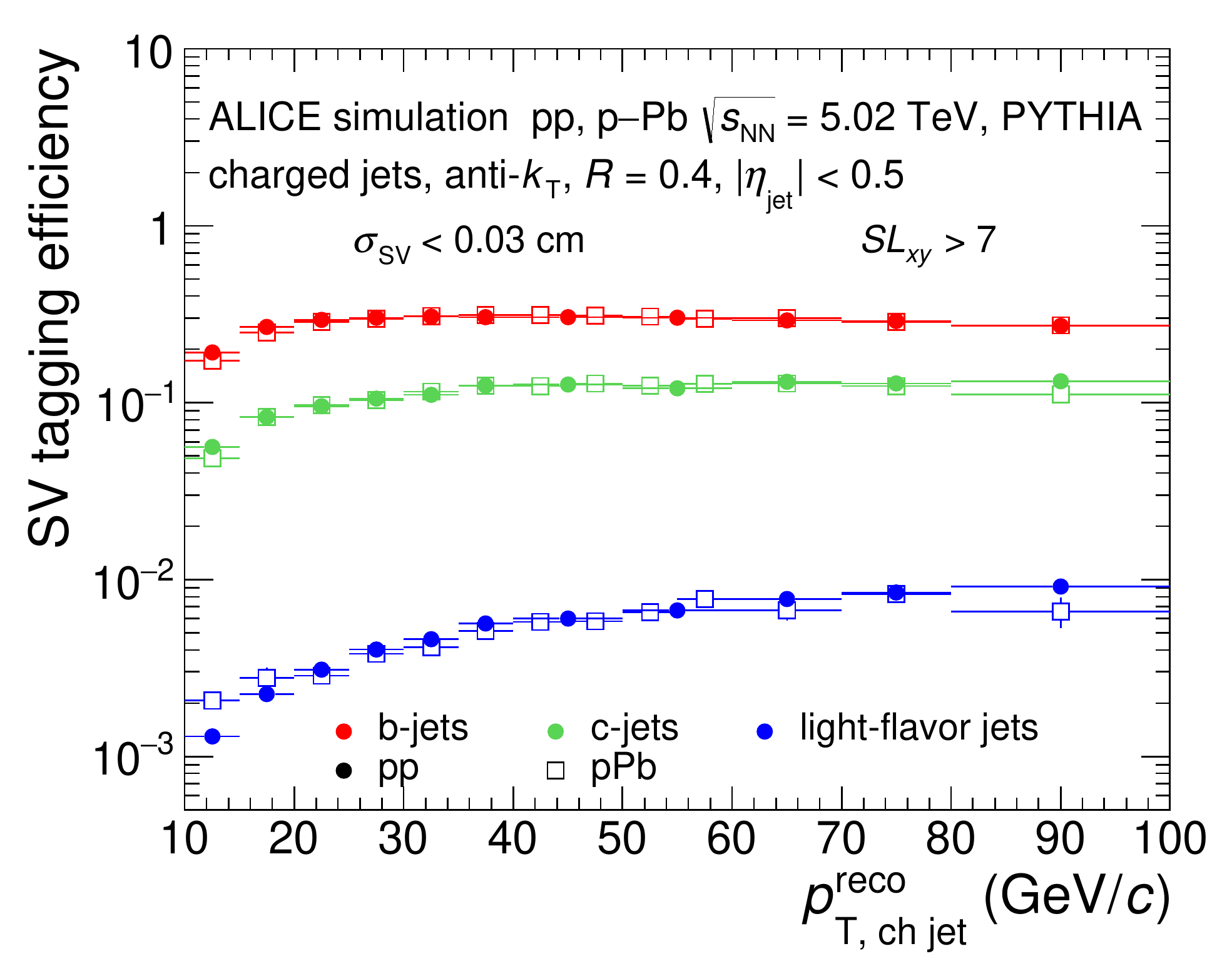}%
\centering%
\caption{Beauty-jet tagging efficiencies, as well as charm-jet, and light-flavor jet mistagging efficiencies for the SV method in pp (solid markers) and \pPb (open markers)  collisions  at $\snn=5.02$\,TeV, shown as a function of jet transverse momentum.}%
\label{fig:TaggingEff_SV}
\end{figure}

For the SV method, tagging and mistagging efficiencies of beauty, charm, and light-flavor jets were estimated based on the same detector-level MC simulation data sets that were used in the IP method. While the IP algorithm used the MC simulation to get templates and assesses the reconstruction efficiency with a data-driven method, the SV algorithm obtained the efficiency directly from the MC simulation via Eq.~(\ref{eq:TaggingEff}). In particle-level simulations, a jet was counted as a b jet if there was a beauty hadron present with a three-momentum vector contained within the jet cone. An analogous definition was also used for c jets and the remaining jets were considered to be light-flavor jets.
Figure~\ref{fig:TaggingEff_SV} presents the efficiencies
as a function of jet momentum in pp and \pPb collisions.
The figure shows that tagging with the default selection criteria yields similar performance in both systems, ensuring suppression of light-flavor jets by two orders of magnitude. Comparing the tagging efficiencies of the IP and SV methods, it can be seen that the efficiency of the IP method tagging is about a factor two higher because of the less stringent selections that are applied.

\subsection{Purity of the b-jet sample}\label{sec:Purity}

The b-jet tagging algorithms introduced in Sec.~\ref{sec:btag} select not only b jets but also a certain fraction of charm and light-flavor jets, cf. Sec.~\ref{sec:TaggingEfficiency}. Given the higher production cross section of light-flavor and charmed jets, this leads to a significant sample contamination that needs to be corrected for. The purity of the tagged sample of b-jet candidates, $P_{\rm b}$, is defined as the fraction of true b jets over the total number of tagged jets,
\begin{equation}
P_{\rm b}(\pTchjet^{\rm reco}) = \frac{N_{\rm b\,jet}^{\rm tagged} (\pTchjet^{\rm reco})}{N^{\rm tagged} (\pTchjet^{\rm reco})}\ .
\end{equation}
Here $N_{\rm b\,jet}^{\rm tagged}$ is the number of tagged true b jets and $N^{\rm tagged}$ is the number of all tagged jets. 
One of the biggest challenges in the b-jet analysis is to obtain an accurate purity estimate. 

\subsubsection{b-jet purity from the IP tagging}

\begin{figure}[h!]%
\centering

{\includegraphics[width=.65\linewidth]{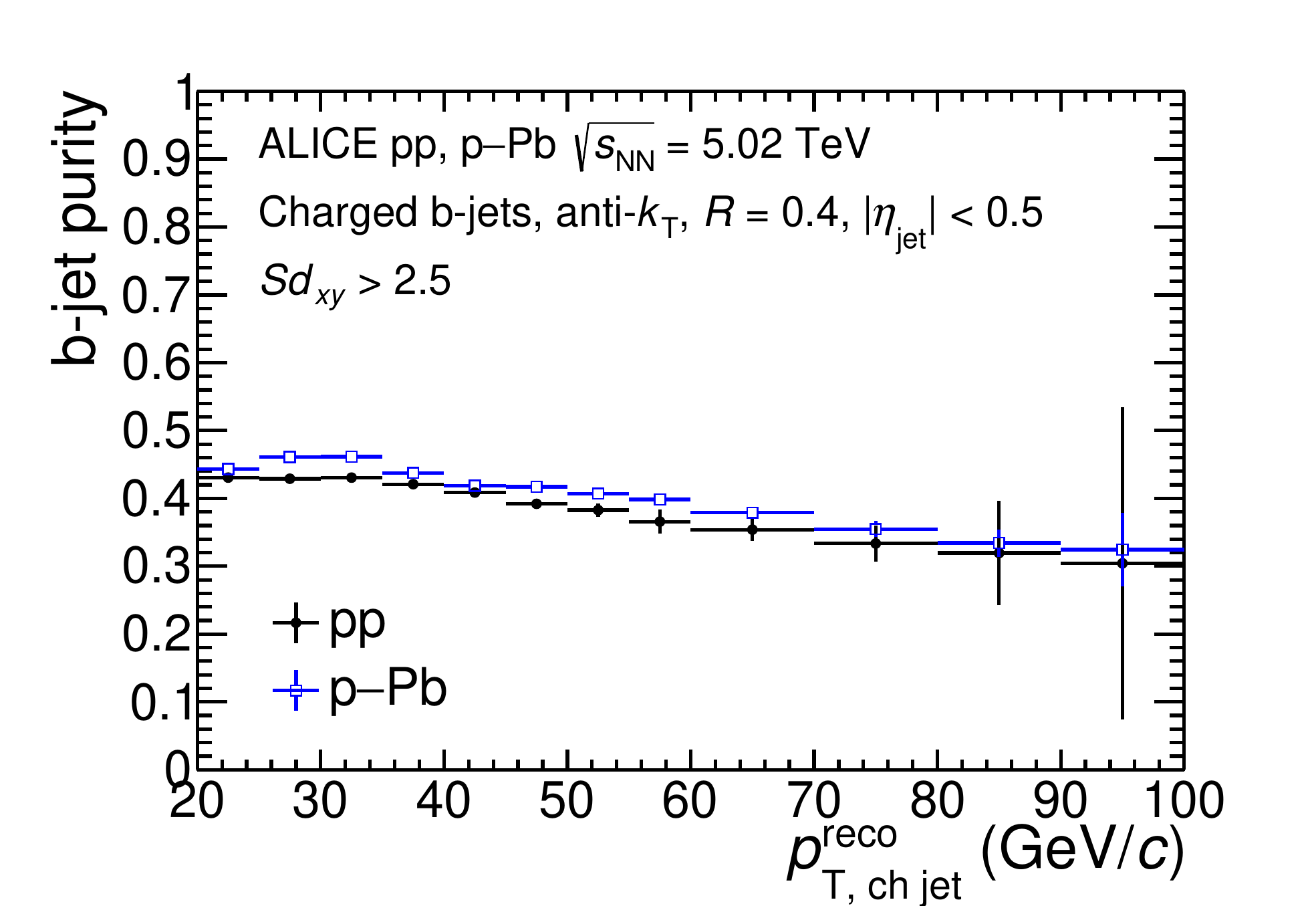}}%

\caption{The b-jet purity as obtained from the IP method in \pp and \pPb collisions.}
\label{fig:bjet_Purity_DATA}
\end{figure}

In the IP method analysis, b-jet purity is estimated using a data-driven method based on the jet probability discriminator. A linear combination of detector-level MC templates corresponding to pure beauty, charm, and light-flavor jets were fitted to the $-\ln(JP)$ distribution measured in data in a similar way as discussed in Sec~\ref{sec:TaggingEfficiencyIP}. Figure~\ref{fig:bjet_Purity_DATA} shows the resulting b-jet purity for the IP method with 
$Sd_{xy}^{\rm min}=2.5$ in \pp and \pPb collisions. 
The template fitting procedure was repeated with other discriminators to assess the corresponding systematic uncertainty, as detailed in Sec~\ref{sec:TaggingEfficiencyIP}.

On figure~\ref{fig:bjet_Purity_DATA}, one can see that the purity in \pPb collisions is slightly higher than that in \pp collisions. This effect arises from small differences between the two systems. Let us note that this difference is much smaller than the systematic uncertainties corresponding to the purity calculation.

\subsubsection{b-jet purity from the SV tagging}
\begin{figure}[h!]
\includegraphics[width=.5\textwidth]{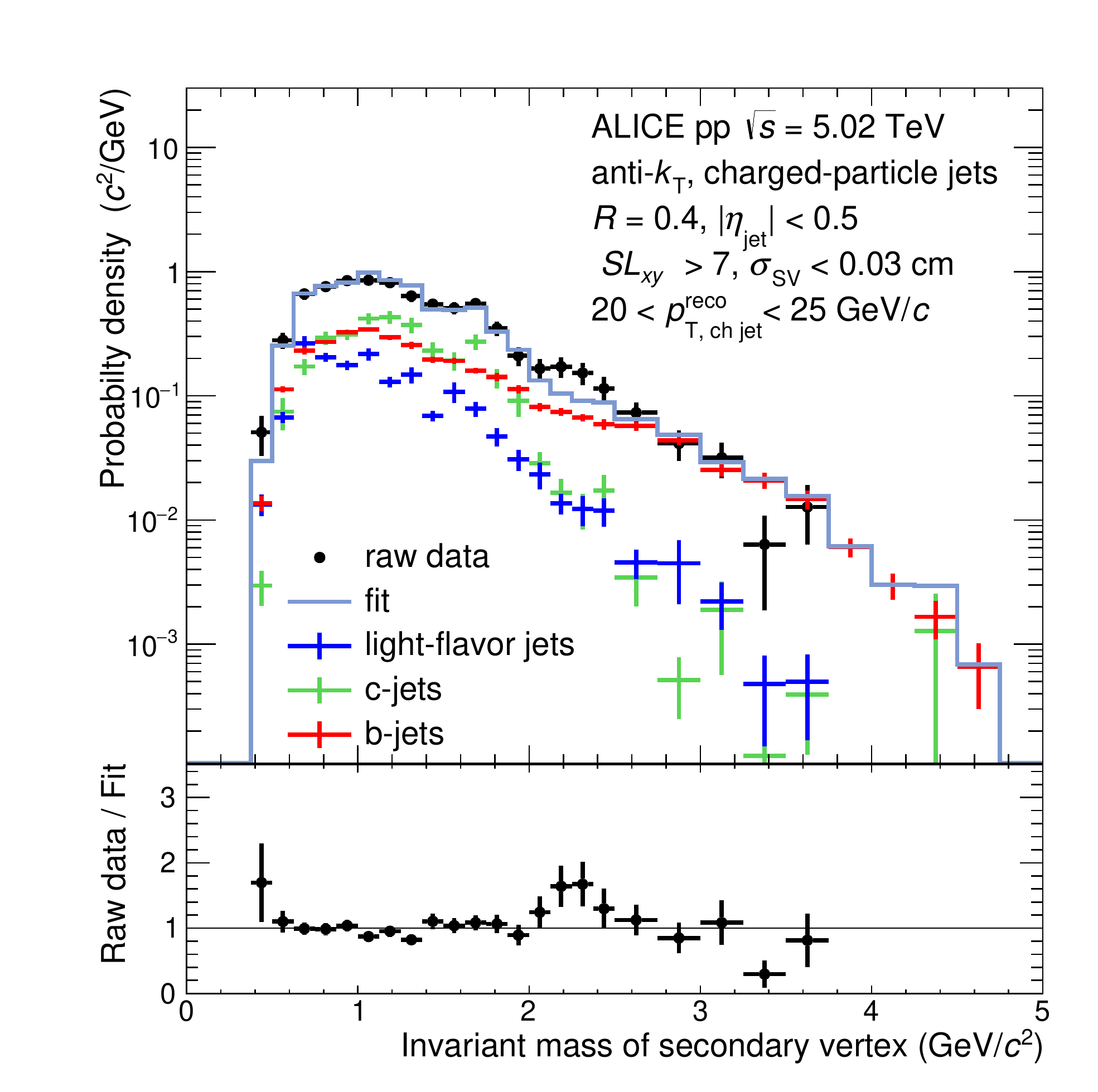}%
\includegraphics[width=.5\textwidth]{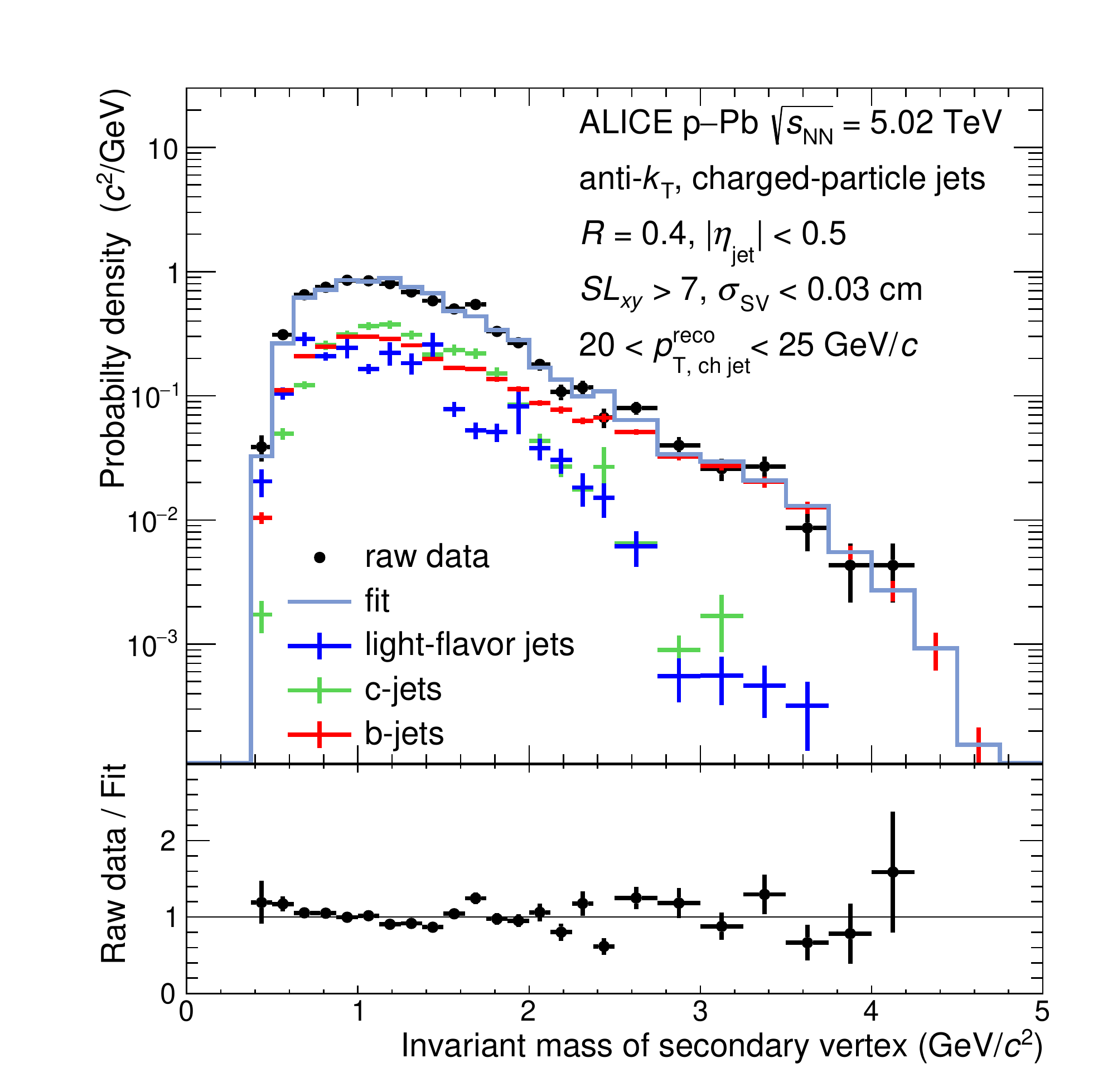}
\centering
\caption{Invariant mass distribution of the combination of three prongs, forming the most displaced
secondary vertex in jets with $20<\pTchjet^{\rm reco}<30$\,\GeVc, tagged with the default selection
$SL_{xy}>7$ and $\sigma_{\rm SV}<0.03$\,cm for \pp (left) and \pPb (right) collisions. The data (black points) are fitted with detector-level MC templates corresponding to beauty, charm, and light-flavor jets to assess
the purity of the b-jet candidate sample. See text for further information on MC.%
} 

\label{fig:puritySVtemplate}
\end{figure}

The purity of the b-jet candidate sample tagged with the SV method was estimated based on a hybrid method that utilizes both data-driven template fitting and simulations. In \pPb data, the purities were primarily determined by fitting the invariant mass distribution of the most displaced secondary vertex with beauty, charm, and light-flavor templates. The invariant mass was calculated from the three prongs that were used to reconstruct the secondary vertex, assuming that all tracks have the mass of a charged pion. These templates were obtained from the detector-level EPOS simulation with embedded PYTHIA~6 events. Analogous fits were done also for the \pp data using detector-level PYTHIA templates. The fits were done in several \pT intervals.
Figure~\ref{fig:puritySVtemplate} shows a typical example of template fit in \pp and \pPb collisions.
The small statistical samples, however, prevented the use of the template fitting method for jets with momenta larger than 30--40\,GeV/$c$.  Therefore, the purity was also estimated based on POWHEG HVQ simulations~\cite{Frixione:2007nw} 
with the CTEQ6M parton distribution function (PDF) set~\cite{Pumplin:2002vw}. In the case of the \pPb system, the EPS09 nPDF set was applied in addition~\cite{Eskola:2009uj}, and the rapidity shift was taken into account.
Simulated particle-level charm and beauty jet \pT spectra were subjected to instrumental (efficiency and detector effects) and background fluctuation effects to estimate the c- and b-jet contributions in the inclusive raw jet spectrum before tagging. The purity was then estimated in each $\pTchjet^{\rm reco}$ bin as
\begin{equation}
    P_{\rm b} = \frac{ \epsilon_{\rm b} N_{\rm b}   }{\epsilon_{\rm b} N_{\rm b} + \epsilon_{\rm c} N_{\rm c} + \epsilon_{\rm lf}\left(N_{\rm incl} - N_{\rm b} - N_{\rm c} \right)}\,,
\end{equation}
where $\epsilon_{\rm b}$,  $\epsilon_{\rm c}$, and  $\epsilon_{\rm lf}$
are tagging and mistagging efficiencies for  beauty, charm, and light-flavor jets, respectively; and $N_{\rm b}$ ($N_{\rm c}$) is the estimated contribution of beauty (charm) jets in the raw inclusive untagged jets $N_{\rm incl}$.
Nevertheless, this purity estimate relies on model parameters that cannot be directly validated, i.e, quark masses as well as renormalization and factorization scales used in the computation of the beauty and the charm production cross section.
Hence, a statistical analysis was carried out comparing simulated purities with purities obtained by the data-driven invariant mass template fit method simultaneously in a broad range of tagging selection criteria. This was done in order to determine the simulation configurations that are consistent with the results of the data-driven method. Consistency was defined with a $\chi^2/{\rm NDF}<10$ goodness-of-description test taking into account the total number-of-degrees-of-freedom (NDF) in the simultaneous comparison.
The configuration space covered variations of the QCD renormalization and factorization scales by factors 0.5--2 with respect to the default values, and variations of the quark masses in the range 4.5--5 \GeVmass\ for b-quarks and 1.3--1.7 \GeVmass\ for c-quarks.
The variation of the heavy quark masses only has a small effect on the observed b-jet sample purity, below 2\% for the b-quark and negligible for the c-quark. Changing the factorization (renormalization) scales in the simulation of the b-quark spectrum by a factor of 2 affects the purity in the same (opposite) direction by 4 to 8\%, while a factor of 2 change in the renormalization or factorization scales in the simulation of the c-quark spectrum causes a 2 to 6\% effect on the resulting purity in either direction.

\begin{figure}[h!]
\includegraphics[width=0.5\textwidth]{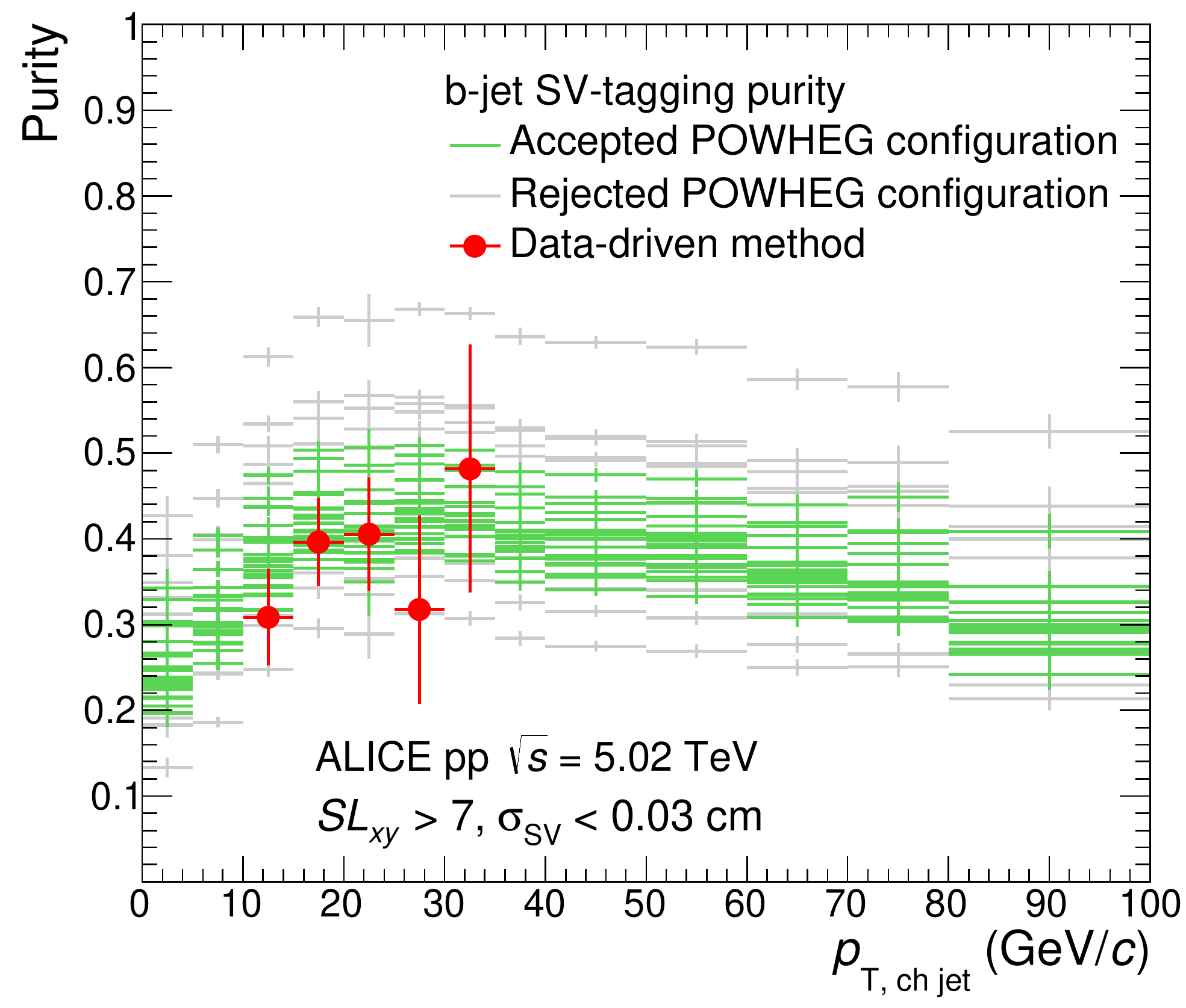}%
\includegraphics[width=0.5\textwidth]{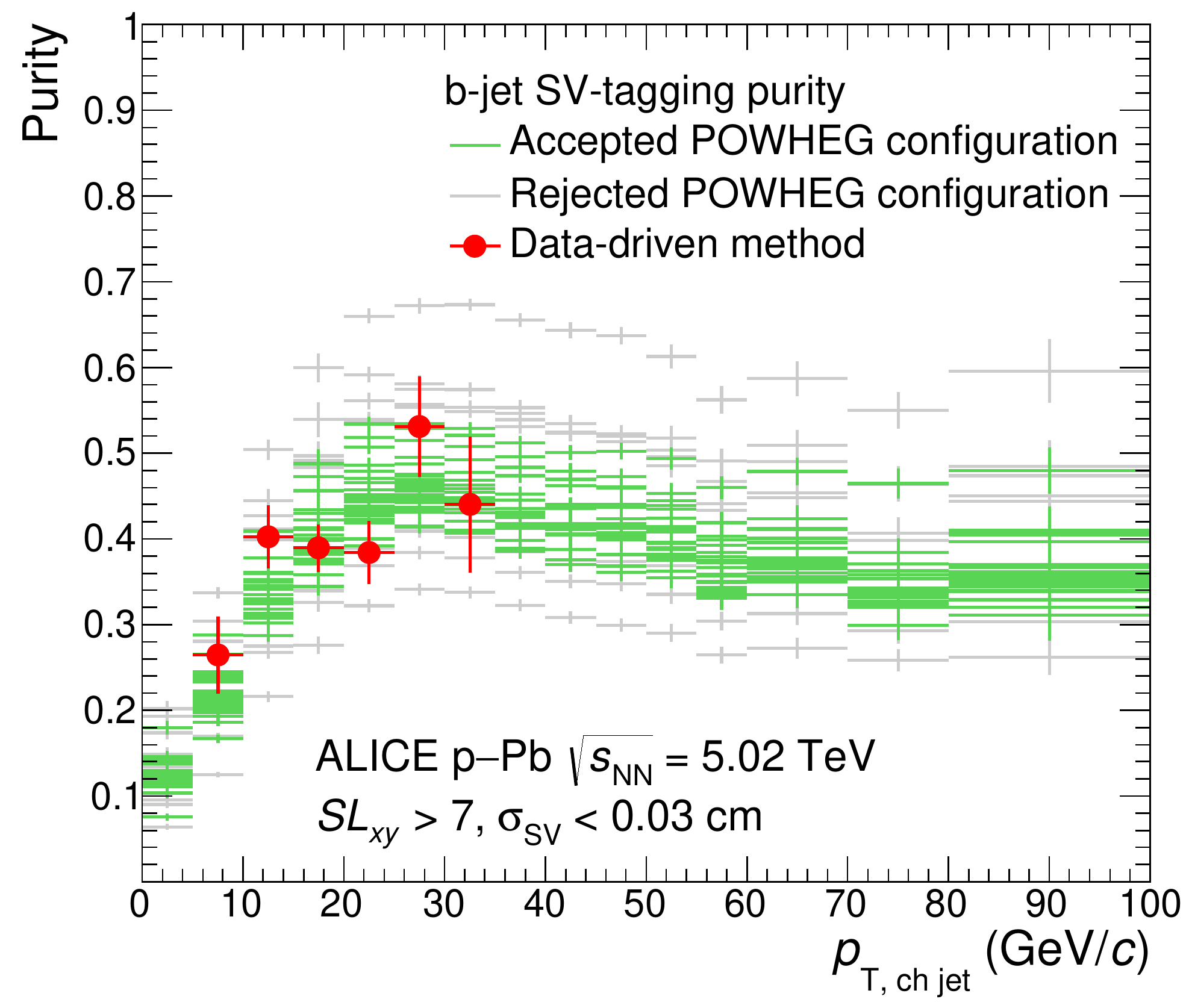}
\centering
\caption{Purity of the b-jet candidates selected with the SV method when using the default tagging selection criteria. The purity was estimated with the data-driven template fit method (red points) and with the POWHEG-simulation based approach. The POWHEG scale variations accepted by the statistical analysis are colored green, the rejected ones are gray. Results for \pp and \pPb  are shown in  the left and  right panel, respectively.}
\label{fig:puritySVhybrid}
\end{figure}

Simulations with accepted configurations were then used to determine the purities in the \pPb as well as the \pp data.
Figure~\ref{fig:puritySVhybrid} shows a comparison of the b-jet sample purity obtained for the default tagging with the template fit method and the POWHEG-simulation-based approach. All accepted configurations were used to assess the systematic uncertainty related to the purity of the tagged b-jet candidate sample.

\subsection{Detector effects and unfolding}
\label{sec:unfold}

The measured jet spectra were affected by distortions stemming from two main sources: instrumental effects and local background fluctuations with respect to the mean underlying event density. These two effects smeared the true jet spectrum and can be corrected for via an unfolding procedure. The corrections were assumed to factorize; thus, were handled with a product of two matrices that were determined separately~\cite{Adam:2015ewa}.
The instrumental effects were accounted for by constructing a response matrix that is based on a b-jet sample generated with PYTHIA~8~\cite{Sjostrand:2006za}, and subsequently processed with an ALICE GEANT~3-based particle transport model~\cite{Brun:1994aa}. The detector-level jets were matched to the particle-level jets based on geometry. This was done by minimizing their angular distance $\Delta R = \sqrt{\Delta\varphi^{2} + \Delta\eta^{2}}$, where $\Delta\varphi$ and $\Delta\eta$ are, respectively, the differences in azimuthal angle and pseudorapidity between given particle-level and detector-level jets.  One-to-one correspondence between particle-level and detector-level jets was required, and $\Delta R$ was constrained to be less than 0.25~\cite{Abelev:2013kqa}.
The instrumental effects cause a similar shift in the jet energy scale of reconstructed charged-particle b jets and inclusive untagged jets; untagged jets being shifted  by about $\approx 1$\% more. The shift is \pT-dependent and for b jets with 10\,\GeVc reaches about 2\% and increases to about 18\% for 100\,\GeVc b jets. The jet energy scale resolution of b jets and inclusive untagged jets is likewise similar; b jets having by about 1\% smaller resolution than the untagged jets. The resolution for 10\,\GeVc b jets is about 17\% and increases to approximately 22\% for 100\,\GeVc b jets.

The matrix that describes momentum smearing due to background fluctuations was obtained with two methods based on track embedding and the random cone technique (RC)~\cite{Abelev:2012ej}.  
In the track-embedding approach, a track was embedded perpendicular in azimuth to the axis of the tagged b-jet candidate. This region is expected to be dominated by the underlying event. The resulting momentum smearing is
\begin{equation}
    \delta\pT^{\rm emb} = \pTchjet^{\rm raw,\,emb} - \rho A_{\rm jet} - \pT^{\rm emb}\,,
\end{equation}
where $\pTchjet^{\rm raw,\,emb}$ is the reconstructed momentum of the jet with the embedded track, $A_{\rm jet}$ is its area, $\rho$ is the estimated underlying event \pT density, and $\pT^{\rm emb}$  is the transverse momentum of the embedded track.
 
In the RC approach, momentum smearing was calculated using a cone with radius $R_{\rm cone}=0.4$ placed in a random position in the $\eta-\varphi$ plane in an event.
This cone must not overlap with the leading and the sub-leading jets in the event and must be fully inside the acceptance of the central barrel.

The momentum smearing is calculated from tracks which are inside the cone as:
\begin{equation}
\delta \pt^{\rm RC} = p_{\rm T}^{\rm RC} \, - \, \rho \pi R^2_{\rm cone}\,,
\end{equation}
where $p_{\rm T}^{\rm RC}$ denotes the sum of the \pT of the tracks inside the cone. Only events which contained a tagged b-jet candidate were selected for the calculation of $\delta\pT$.

The $\delta\pt$ matrices obtained with the track embedding and RC techniques provided consistent unfolded b-jet spectra and the difference is accounted for as a systematic uncertainty. 
In this analysis, the track embedding technique was used in the standard analysis, and the RC method as a systematic variation.

By default, unfolding of the raw b-jet spectrum defined in Eq.~(\ref{eq:rawspeccorr}) was performed using the singular value decomposition (SVD) method~\cite{Hocker:1995kb} implemented in the RooUnfold package~\cite{Adye:2011gm}.
The optimal regularization parameter value was found to be four for the SV analysis and eight for the IP analysis. Stability of the unfolded solutions was tested also with the Bayesian unfolding~\cite{2003RPPh...66.1383D} and the $\chi^2$ unfolding.  These algorithms provided consistent results with the SVD and the differences were taken into account in the systematic uncertainties.

\subsection{b-jet cross section and nuclear modification factor  }

The \pt-differential b-jet production cross section was calculated as
\begin{equation}
\frac{\mathrm{d}^{2} \sigma ^{\rm b\,\,jet}}{\mathrm{d} \pTchjet \mathrm{d}\eta_{\rm jet}} = \frac{1}{\mathcal{L}} \times 
\frac{{\rm d}^{2} N^{\rm b\,\,jet}_{\rm unfolded}}{{\rm d} \pTchjet {\rm d}\eta_{\rm jet}}\,,
\label{eq:bjet_xsec}
\end{equation}
where ${\rm d}^{2} N^{\rm b\,\,jet}_{\rm unfolded}/{\rm d}\pTchjet {\rm d}\eta_{\rm jet}$ is the unfolded \pt differential yield of b jets and $\mathcal{L}$ is the integrated luminosity corresponding to minimum bias events, which was quoted for the \pp and \pPb data samples in Sec.~\ref{sec:DATA}.

Modification of the b-jet spectrum in \pPb collisions due to nuclear matter effects was then quantified with the nuclear modification factor~\cite{Miller:2007gl}, which compares the \pT-dependent production rates in \pPb to the rates expected from the independent superposition of pp collisions.
\begin{equation}
\RpPbBjet = \frac{1}{A} \frac{\mathrm{d}^{2}  \sigma _{\rm pPb}^{\rm b\,\,jet} / \mathrm{d} \pTchjet \mathrm{d}\eta_{\rm jet} }{ \mathrm{d}^{2}  \sigma _{\rm pp}^{\rm b\,\,jet} / \mathrm{d} \pTchjet \mathrm{d}\eta_{\rm jet}  }\ ,
\label{eq:bjet_RpPb}
\end{equation}
where $A=208$ is the number of nucleons in the Pb nucleus.

\subsection{Combining the results of the IP and SV methods}\label{sec:BLUE_merging}

The \pt-differential b-jet production cross sections obtained from the IP and SV methods were combined using the Best Linear Unbiased Estimator (BLUE) method~\cite{Lyons:1988rp,Valassi:2003mu}. The BLUE method is used to combine different measurements of the same physical quantity, where the uncertainties of the individual measurements are correlated between the measurements to some extent. Besides the b-jet cross section in \pp and \pPb collisions, the BLUE method was also used to obtain the combined nuclear modification factor \RpPbBjet 
given that correlated systematic uncertainties cancel to a different degree in the individual ratios for the IP and SV analyses.

The combined results were obtained under the following considerations. The systematic uncertainties from tagging, and purity extraction were assumed to be uncorrelated between the two methods. The contributions to the systematic uncertainty from the tracking efficiency and \pT\ resolution, as well as from the  contamination by secondary tracks, were treated as fully correlated. 
Since the same data set was used in the two methods, the statistical uncertainty is partially correlated. The correlation coefficient $\rho_{\rm stat}$ was estimated as
\begin{equation}
    \rho_{\rm stat} = \frac{{\rm Cov}({\rm IP},{\rm SV})}{\sigma_{\rm IP} \sigma_{\rm SV} }   \;\;\; \mbox{with} \;\;\;  {\rm Cov}({\rm IP},{\rm SV}) = \frac{\sigma_{\rm IP}^{2}  \sigma_{\rm SV}^{2}}{ \sigma_{{\rm IP} \cap {\rm SV}}^{2}}\,, 
\end{equation}
where $\sigma_{\rm IP}$ ($\sigma_{\rm SV}$) is the statistical uncertainty corresponding to the jet sample from the IP (SV) method, and $\sigma_{{\rm IP} \cap {\rm SV}}$ is the statistical uncertainty corresponding to the sample selected by both the IP and the SV methods.
The correlation coefficients for statistical uncertainty are $\rho_{\rm stat}=0.35$ for \pp collisions and $\rho_{\rm stat}=0.27$ for \pPb collisions.
For the background fluctuations and unfolding uncertainties, which were partially correlated between both methods, an arbitrarily chosen correlation coefficient value of 0.5  was used, with values 0 and 1 used as consistency checks. Correlation coefficients between other parameters were varied similarly and the resulting systematic uncertainty from these choices was found to be negligible.

\section{Sources of systematic uncertainties}
\label{sec:Systematic}
Systematic uncertainties of the \pt-differential b-jet cross section and \RpPbBjet were assessed by varying the {selection and} correction procedures. Table~\ref{tab:ParameterVariationsSV} lists the possible sources of systematic uncertainties, and the adopted variations, with respect to the standard selection procedures and methods used to obtain the central values of the results. These variations are discussed in more detail below. Table~\ref{tab:SysTotal} provides a summary of all uncertainties, reported separately for the IP and SV analyses, as well as for the combined results obtained with the BLUE method.
The two analyses were developed largely independently from each other.
In the IP analysis, all uncertainties were considered as symmetrical, while in the SV analysis, most of the uncertainties were considered as asymmetrical. Systematic uncertainties due to the tracking efficiency and \pT resolution, tagging, contamination by secondary tracks, and background fluctuations were treated as correlated between the \pp and \pPb systems. Hence, these were partially propagated into \RpPbBjet, taking the correlation into account. All the other uncertainties were considered uncorrelated and were fully propagated. The different types of correlated systematic uncertainties on the \RpPbBjet were determined by simultaneously varying the \pp and the \pPb results to make sure that the correlations cancel out. Since the combination with the BLUE method requires symmetric uncertainties, two SV spectra were made, one with the lower and one with the upper uncertainties. These spectra were combined with the IP spectrum separately, and a conservative choice was made by taking the maximum of the lower and upper boundaries point-by-point in the combined result.
The individual uncertainty sources are discussed in detail in the following paragraphs.

\begin{table}[htbp]
\caption{Summary of sources of systematic uncertainty and adopted variations to estimate their effects in the cases of the SV and the IP methods.}
\begin{center}
\setlength\tabcolsep{3pt} 
\begin{tabular}{|l|l|l|c|c|}
\hline                                         
 \multicolumn{3}{|c|}{Source}  & Standard analysis & Variations\\
\hline\hline
\multirow{1}{*}{\rotatebox{90}{Common}} 
& \multicolumn{2}{l|}{Tracking efficiency} & default reconstruction & 4\% of tracks removed \\
& \multicolumn{2}{l|}{Track \pT resolution} & default reconstruction & \pT-smeared reconstruction \\
& \multicolumn{2}{l|}{Secondary track contamination} & default MC correction & data-driven estimate \\
& \multicolumn{2}{l|}{Underlying event fluctuations} & embedding & random cone \\

\hline\hline
\multirow{1}{*}{\rotatebox{90}{IP method\phantom{...........}}} 
&  \multirow{5}{*}{Unfolding}  & Method & SVD & Bayesian--$\chi^2$ \\
&            & Regularization & 8 & 7--9 \\
&            & Matrix & full & truncated at 5 GeV/$c$ \\
&            & Binning & default & limits shifted by 2 GeV/$c$ \\
&            & Prior function & POWHEG b-jet spectrum & measured--unfolded $\chi^2$ spectra \\
\cline{2-5}
& \multirow{2}{*}{Tagging efficiency/purity} & $Sd_{xy}$& 2.5        & 1--4 \\
&              & Fit distribution & $-\ln(JP)$ & $f_E$ \\
\cline{2-5}

\hline\hline
\multirow{1}{*}{\rotatebox{90}{SV method\phantom{.............}}} 
& \multirow{2}{*}{Tagging efficiency} & $SL_{xy}$& 7        & 6--8 \\
&              & $\sigma _{\rm SV}$   & 0.03\,cm & 0.02--0.05\,cm \\
\cline{2-5}
& \multicolumn{2}{l|}{Purity} & POWHEG b and c spectra & ``hybrid'' scale variations \\
\cline{2-5}
&  \multirow{5}{*}{Unfolding}  & Method & SVD & Bayesian \\
&            & Regularization & 4 & 3--5 \\
&            & Matrix & full & truncated at 5 GeV/$c$ \\
&            & Binning & default & limits shifted by 1 GeV/$c$ \\
&            & Prior function & POWHEG b-jet spectrum & all scale variations \\

\hline
\end{tabular}
\end{center}
\label{tab:ParameterVariationsSV}
\end{table}

{\small
\begin{table}[h!]
\caption {Statistical and systematic uncertainties, in percent, corresponding to three representative \pTchjet ranges for the \pp and \pPb cross sections, as well as for the \RpPbBjet. Uncertainties of the IP and SV methods are quoted separately. Wherever applicable, the table also reports the resulting combined uncertainties. Both the upper and lower values are listed for the asymmetric SV systematic uncertainties. An additional uncertainty from the normalization by the integrated luminosity~\cite{ALICE-PUBLIC-2016-005,Abelev:2014epa} is quoted in the last row.
} 
  \begin{center}
    \setlength\tabcolsep{3pt} 
    \begin{tabular}{|l|l||c||c|c|c||c|c|c|} 
    \hline
    \multicolumn{2}{|c||}{\pTchjet interval} & \multicolumn{1}{c||}{ 10--20\,\GeVc} & \multicolumn{3}{c||}{  40--50\,\GeVc} & \multicolumn{3}{c|}{ 80--100\,\GeVc} \\
    \hline
    \multicolumn{2}{|c||}{analysis} & SV & IP & SV & comb. & IP & SV & comb. \\
    \hline\hline
    \multirow{3}{*}{\shortstack[l]{Statistical\\uncertainty}}   
        &  \pp  & 1.9 & 6.3 & 5.2 & 6.3 & 23.5 & 18.4 & 22.9 \\
        &  \pPb & 1.9 & 3.7 & 3.6 & 3.8 & 9.0 & 12.9 & 9.3 \\
        & \RpPbBjet & 2.6 & 8.9 & 6.4 & 5.8 & 31.2 & 22.5 & 20.6 \\
    \hline\hline
    \multirow{3}{*}{\shortstack[l]{Tracking\\efficiency}}
        &  \pp  & 7.9 & 11.1 & 8.4 & 11.1 & 16.2 & 9.5 & 16.0 \\
        &  \pPb & 6.7 & 12.1 & 9.2 & 12.2 & 14.2 & 8.6 & 14.5 \\
        & \RpPbBjet  & 1.4 & 0.9 & 1.0 & 1.0 & 1.9 & 1.3 & 1.6 \\
    \hline
    \multirow{3}{*}{\shortstack[l]{Tracking\\resolution}}
        &  \pp  & $+1.2/-1.2$ & 1.4 & $+3.9/-3.9$ & 1.4 & 3.1 & $+6.0/-6.0$ & 3.2 \\
        &  \pPb & $+3.3/-3.3$ & 1.6 & $+4.5/-4.5$ & 1.5 & 2.1 & $+5.3/-5.3$ & 1.9 \\
        & \RpPbBjet & $+2.1/-2.2$ & 0.2 & $+0.6/-0.6$ & 0.5 & 1.0 & $+0.8/-0.7$ & 0.9 \\
    \hline
    \multirow{3}{*}{\shortstack[l]{Secondary vertex\\contamination}}
        &  \pp  & $+1.6/-0.0$ & 2.3 & $+2.4/-0.0$ & 2.6 & 4.0 & $+2.9/-0.0$ & 7.2 \\
        &  \pPb & $+4.1/-0.0$ & 6.1 & $+5.4/-0.0$ & 1.5 & 7.6 & $+7.8/-0.0$ & 2.2 \\
        & \RpPbBjet  & $+0.0/-2.5$ & 3.6 & $+0.0/-3.0$ & 1.0 & 3.3 & $+0.0/-5.1$ & 4.2 \\
    \hline
    \multirow{3}{*}{\shortstack[l]{Background\\fluctuation}}
        &  \pp  & $+0.0/-5.4$ & 6.9 & $+0.0/-10.3$ & 6.8 & 3.2 & $+3.2/-0.0$ & 3.1 \\
        &  \pPb & $+0.0/-3.1$ & 2.7 & $+0.0/-5.4$ & 2.8 & 1.7 & $+3.1/-0.0$ & 1.8 \\
        & \RpPbBjet & $+2.6/-0.0$ & 4.2 & $+6.7/-0.0$ & 5.5 & 1.4 & $+1.4/-0.0$ & 0.7 \\
    \hline
    \multirow{3}{*}{b-jet tagging}
        &  \pp  & $+0.9/-2.8$ & 0.2 & $+3.4/-6.5$ & 0.2 & 3.4 & $+6.8/-13.4$ & 3.3 \\
        &  \pPb & $+3.4/-1.6$ & 0.4 & $+4.6/-8.6$ & 0.5 & 0.8 & $+6.0/-15.3$ & 1.2 \\
        & \RpPbBjet & $+2.5/-2.2$ & 0.4 & $+5.0/-5.3$ & 3.6 & 3.5 & $+10.7/-13.8$ & 7.2 \\
    \hline
    \multirow{3}{*}{Purity}
        &  \pp  & $+13.0/-21.8$ & 15.3 & $+16.4/-16.8$ & 12.0 & 15.3 & $+21.8/-17.3$ & 11.7 \\
        &  \pPb  & $+13.1/-21.0$ & 8.9 & $+11.9/-16.3$ & 9.4 & 8.9 & $+21.1/-15.7$ & 9.5 \\
        & \RpPbBjet & $+5.2/-9.4$ & 14.7 & $+5.2/-6.6$ & 5.6 & 14.7 & $+8.2/-9.7$ & 8.8 \\
    \hline
    \multirow{3}{*}{Unfolding}
        & \pp   & $+7.2/-0.9$ & 2.1 & $+1.0/-1.9$ & 2.1 & 7.9 & $+27.1/-6.1$ & 7.8 \\
        &  \pPb & $+9.5/-5.6$ & 0.9 & $+0.5/-4.5$ & 0.9 & 1.6 & $+11.4/-14.3$ & 1.5 \\
        & \RpPbBjet & $+2.5/-5.4$ & 2.2 & $+3.1/-4.7$ & 2.6 & 8.1 & $+4.4/-15.9$ & 10.6 \\
    \hline\hline
    \multirow{3}{*}{\shortstack[l]{Total systematic\\uncertainty}}
        &  \pp  & $+17.0/-24.0$ & 18.1 & $+19.3/-22.8$ & 18.1 & 22.8 & $+37.4/-25.4$ & 23.1 \\ 
        &  \pPb & $+18.6/-23.2$ & 16.5 & $+17.2/-22.2$ & 15.8 & 18.7 & $+28.0/-28.1$ & 17.7 \\ 
        & \RpPbBjet & $+7.3/-11.6$ & 15.9 & $+10.4/-10.2$ & 9.1 & 17.7 & $+14.3/-23.8$ & 16.2 \\

    \hline\hline
    \multirow{3}{*}{\shortstack[l]{Normalization\\uncertainty}}
        &  \pp  & \multicolumn{7}{c|}{ 2.34 } \\
        &  \pPb & \multicolumn{7}{c|}{ 3.7 } \\
        & \RpPbBjet & \multicolumn{7}{c|}{ 4.37 } \\
    \hline
    \end{tabular}
\label{tab:SysTotal}
  \end{center}
\end{table}
}

\subsection{Tracking efficiency}
\label{sec:TrkEff}
The systematic uncertainty on tracking efficiency is about 4\%~\cite{Acharya:2019jyg}. This uncertainty translates into an uncertainty on the energy scale of reconstructed jets. The resulting effect on the b-jet spectra was estimated by constructing an instrumental response matrix from which 4\% of tracks were randomly removed. This matrix represents the downward uncertainty on the reconstruction efficiency. It is assumed that a 4\% variation towards higher tracking efficiency would affect the results symmetrically.
The tracking efficiency uncertainty is one of the major sources of systematic uncertainties on the b-jet cross section. It tends to increase with increasing b-jet \pt.

\subsection{\pT resolution of tracks}
The \pT resolution of tracks was discussed briefly in Sec.~\ref{sec:btag} and more details can be found in Ref.~\cite{Alicedet:ijmpa}.
The systematic uncertainty on track transverse momentum resolution was estimated from the azimuthal variation of the \pT spectrum of positively and negatively charged particles following the procedure described in Ref.~\cite{Acharya:2018jqh}. The resulting effect of these variations on the b-jet cross section spectra was investigated by unfolding the b-jet spectrum with an instrumental response matrix that reflected the observed local variations in track \pT\ smearing.

\subsection{Contamination from secondary tracks}
Contamination of jets from secondary tracks, due to weak decays, was corrected for using the instrumental matrix. This correction is MC based and relies on the secondary track fractions from the simulations. As a systematic variation, these fractions were taken from a data-driven approach where DCA distributions of tracks to the primary vertex were fitted with templates corresponding to primary tracks and secondary tracks. This resulted in a systematic shift in jet energy scale.
In the SV analysis this uncertainty was treated as one-sided since the true fraction of secondary tracks is expected to fall between the two calculations.

The production of long-lived strange particles is known to be poorly described by the PYTHIA MC event generator~\cite{ALICE:2020jsh, Aamodt:2011lig, Chatrchyan:2012lig, Abelev:2012lig}. Decays of \kzero and strange baryons were however found to contribute by less than 1\% to the constructed light-flavor SV invariant mass templates that are used for the data driven purity estimate. Possible variations of the strangeness in simulations would; therefore, have negligible impact on the shape of this template and should have negligible impact on the extracted purity. A similar situation holds also for the IP templates, where decays of long-lived strange particles contribute on the percent level only. Omission of the long-lived strange particles from construction of the templates led to negligible changes of the fit results.

\subsection{Underlying event fluctuations}
This uncertainty was estimated by comparing the spectra unfolded using $\delta\pT$ matrices constructed with the track embedding and the random cone methods. This resulted in a one-sided uncertainty on the SV spectra.

\subsection{b-jet tagging efficiency and purity in the IP method}
The uncertainty was estimated by varying the default impact parameter significance and template fit discriminator. The working point of the tagging selection criterion, set by default as $Sd_{xy}^{\rm min} = 2.5$, was varied in the range from 1 to 4.
This resulted in variations in the data driven b-jet tagging efficiency and purity that were propagated to the b-jet cross sections.

Similarly, The energy fraction carried by charged tracks associated to the secondary jet vertex, $f_E$ was used as template fitting discriminator. 
Differences between these methods were added up in quadrature with the uncertainties from the fitting method to establish the overall uncertainty on the template fitting.

The purity of the selected b-jet candidates can be in principle affected also by the admixture of the long-lived strange V$^0$ particles (K$^0_{\rm s}$ and $\Lambda$/$\overline{\Lambda}$), which result in decay-daughter tracks with large impact parameters.
The possible effect of these daughter tracks on the purity and efficiency of the IP tagging was tested by ignoring those tracks that, when combined with other tracks of the same event, yield an invariant mass compatible with the K$^0_{\rm s}$ or $\Lambda$ hypothesis.
The corresponding systematic effect on the resulting b-jet spectrum was found to be negligible.

\subsection{b-jet tagging efficiency and purity in the SV method}
The default tagging selection, $SL_{xy} > 7$ and $\sigma_{\rm SV} < 0.03$\,cm, was chosen to fall into a region where the simulation adequately describes the data. 
The variations were performed such that one parameter was kept at its default value while the other parameter was altered. In this study, $SL_{xy}$ was varied from 6 to 8, and $\sigma_{\rm SV}$ was varied from 0.02 to 0.05\,cm. Since these two parameters are correlated, the envelope of the systematic variations was considered, constructed using the point-by-point maximal upper and lower variations.

In the SV method, the major source of systematic uncertainty on the b-jet cross sections stems from the purity assessment of the tagged b-jet candidate sample. The uncertainty 
was evaluated by repeating the analysis with each of the accepted POWHEG purity curves shown in Fig.~\ref{fig:puritySVhybrid}.
The uncertainty is defined by the envelope of the resulting spectrum variations. 

The POWHEG configurations that provide a statistically acceptable description of the purity are determined based on template fits in the \pPb system. Since the same configurations are used in the \pp system, the assumptions on the CNM effects in POWHEG \pPb simulations will, counter-intuitively, affect the purity estimation in the \pp system. This effect was estimated based on the comparison of the POWHEG simulations to the existing heavy-flavor \RpPb measurements ~\cite{Acharya:2019hao,Acharya:2019jqD}. This additional, 
independent uncertainty on the SV-method purity in the \pp system was found to be a few percents at low \pt and is vanishing towards higher \pt.

In the SV method, since a three-prong secondary vertex is required and the purity is determined based on template fitting of the invariant mass distribution, a possible incorrect modelling of V$^0$ particles poses negligible impact on the purity.

\subsection{Unfolding}
\label{sec:UnfoSys}
Both the IP and SV methods use SVD unfolding in the standard analysis. To establish the uncertainty stemming from the choice of the unfolding method, the spectra were also unfolded with the Bayesian method, and in the IP analysis, with the $\chi^2$ method in addition. The sensitivity to the choice of regularization parameter was investigated by changing its value within $\pm1$. The unfolding was also repeated with a modified lower \pT limit of the input spectrum from $\pT=5$ GeV/$c$ to $\pT=1$ GeV/$c$. The SV analysis also considered a different input \pT spectrum binning. Both methods used the b-jet POWHEG spectrum as the default prior function in the respective standard analyses. In the IP analysis, the unfolding was repeated using the measured, as well as  the $\chi^2$-unfolded spectra as priors. In the SV analysis, the unfolding was repeated by taking as priors the POWHEG b-jet spectra resulting from different scale and mass variations. The root mean square (RMS) of the differences between these variations and the standard analysis spectra was taken as uncertainty in the IP analysis. In the SV analysis, the statistical and systematic parts were separated using pseudo-experiments with randomized input spectra. The pseudo-experiments were carried out for the standard analysis configuration as well as for each systematic variation. The maximum deviations at each \pTchjet value were taken as asymmetric uncertainties. 

An additional systematic uncertainty stems from the limited knowledge of very low momentum jet production, determined from PYTHIA simulations when constructing the response matrix. This was estimated by using a matrix that was truncated below $\pTchjet=5$\,\GeVc, and the resulting deviation with respect to the standard analysis spectrum was added up in quadrature to the total uncertainty.

\subsection{Normalization}

There are also uncertainties on the normalization of the differential cross section which will be propagated to the nuclear modification factor. The normalization uncertainties are discussed in detail in Ref.~\cite{Abelev:2014epa} for \pp collisions and in Ref.~\cite{ALICE-PUBLIC-2016-005} for \pPb collisions.

\section{Results and discussion}\label{sec:Results}

\subsection{b-jet production cross section in \pp and \pPb collisions}
\label{sec:bjet_xsec}

\begin{figure}[h!]%
\centering
\includegraphics[width=.5\textwidth]{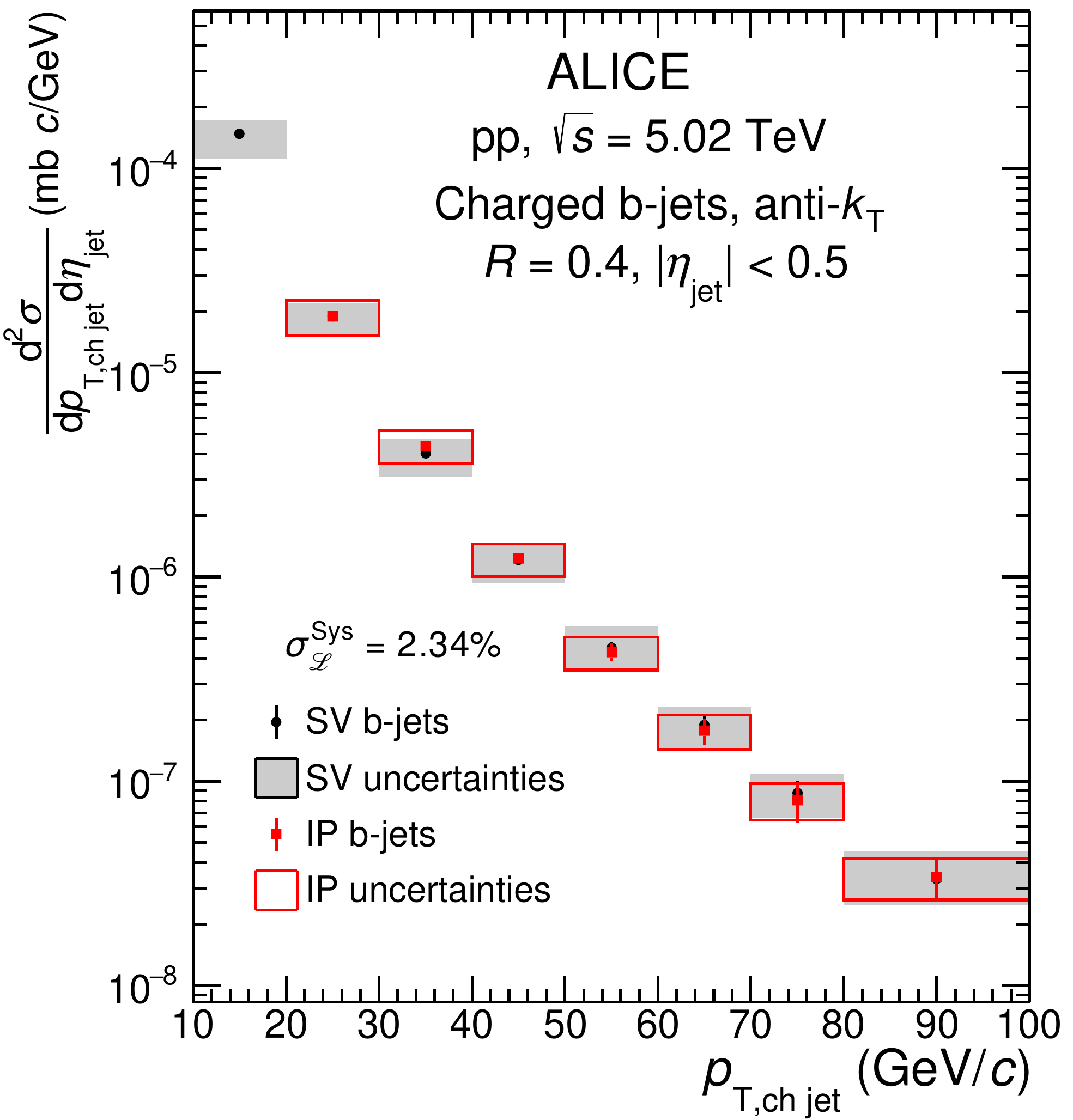}%
\includegraphics[width=.5\textwidth]{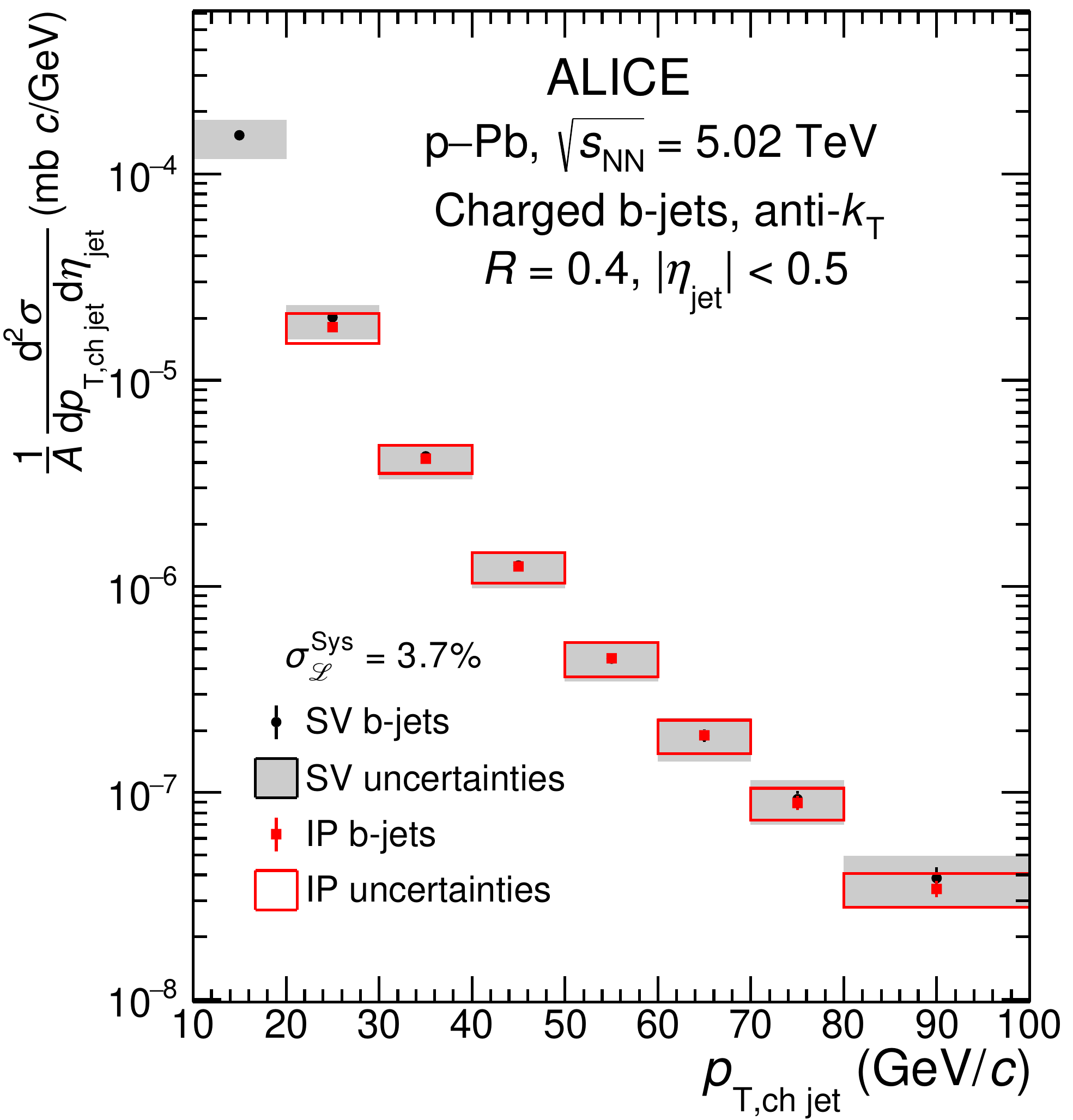}
\caption{Comparison of the \pT differential production cross section of charged-particle anti-$k_{\rm T}$ $R=0.4$ b jets measured in pp and \pPb collisions at $\snn=5.02$\,TeV using the IP and  SV methods. Systematic and statistical uncertainties are shown as boxes and error bars respectively. The additional common normalization uncertainty  due to luminosity  is  denoted $\sigma_{\mathcal{L}}^{\rm Sys}$ and it is quoted separately.}
\label{fig:bjet_pPb_Second}
\end{figure}

\begin{figure}[h!]%
\centering
\includegraphics[width=.5\textwidth]{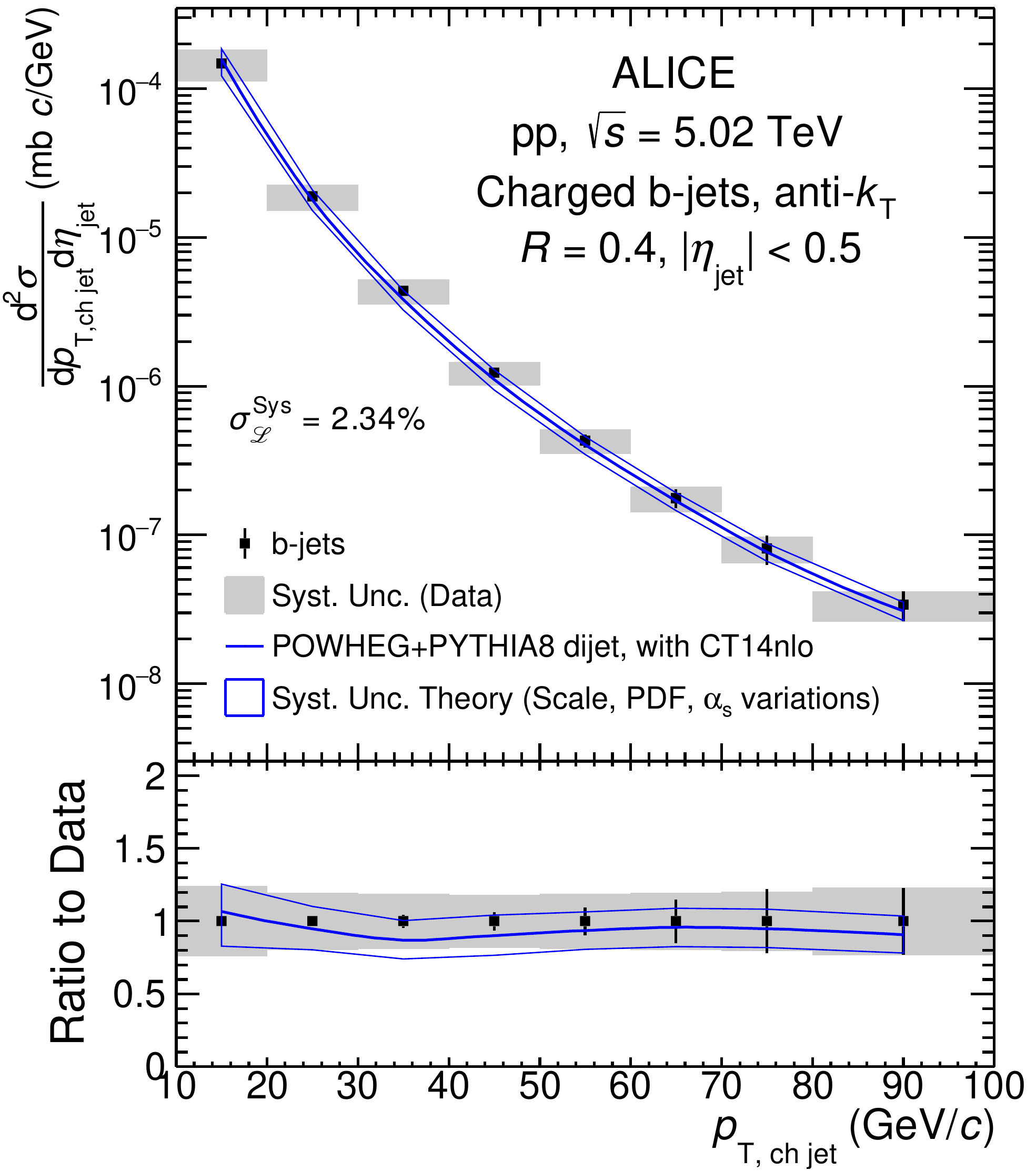}%
\includegraphics[width=.5\textwidth]{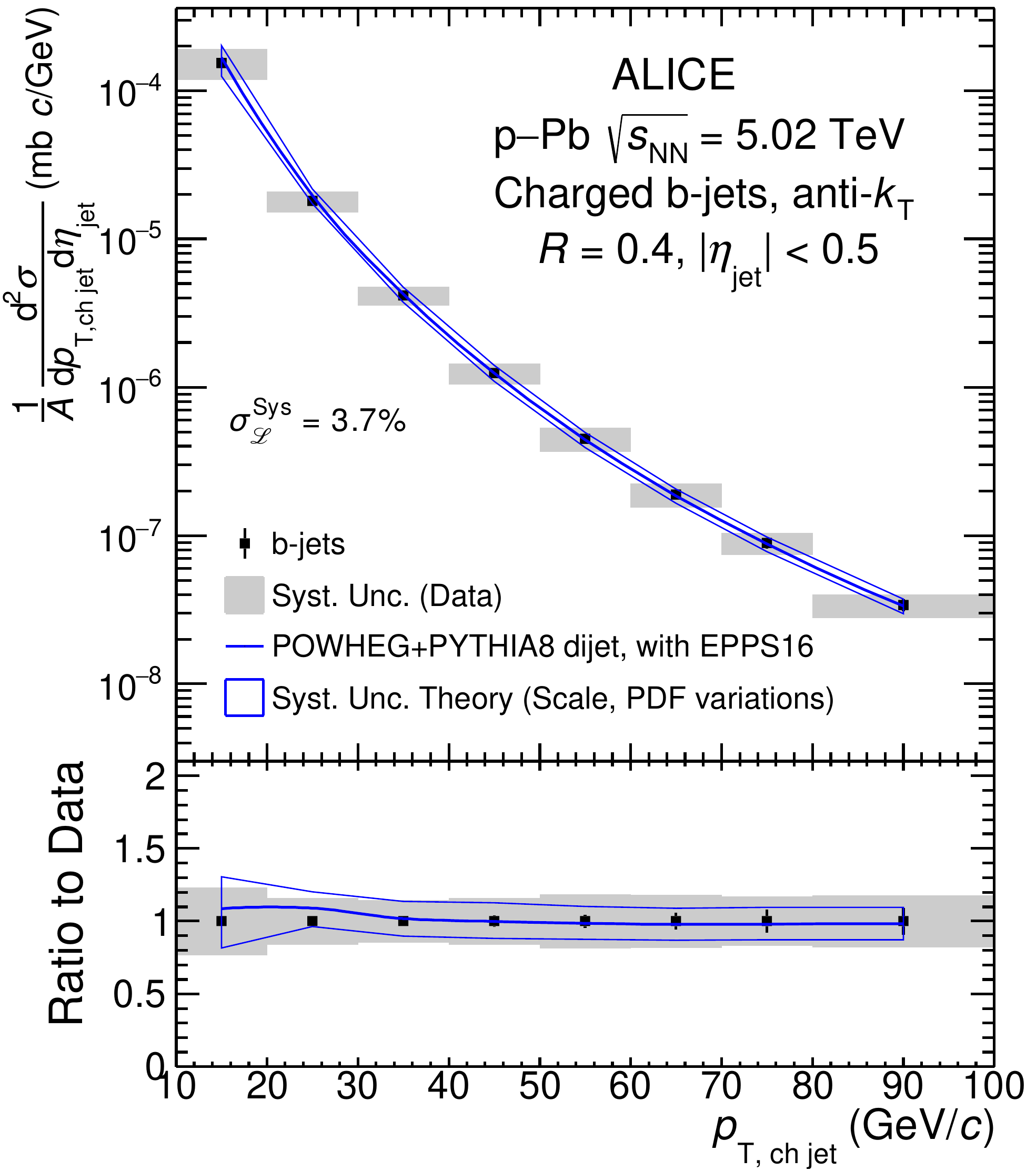}
\caption{Top panels: The combined differential production cross section of charged-particle anti-$k_{\rm T}$ $R=0.4$ b jets measured in \pp (left) and \pPb (right) collisions at $\snn=5.02$\,TeV. The data are compared with a NLO pQCD prediction by the POWHEG dijet tune with PYTHIA~8 fragmentation~\cite{Alioli:2010xa,Frixione:2007vw}. Systematic and statistical uncertainties are shown as boxes and error bars, respectively. The additional common normalization uncertainty  due to luminosity, $\sigma_{\mathcal{L}}^{\rm Sys}$, is quoted separately. Bottom panels: Ratio of the theory calculations to the data.}
\label{fig:bjet_xsec_model}
\end{figure}

Figure~\ref{fig:bjet_pPb_Second} presents the \pt-differential production cross section of b jets
obtained from the IP and SV analyses in pp and \pPb collisions at $\snn=5.02$\,TeV.
For easier comparison across the two systems, the \pPb cross section  is normalized by the number of Pb nucleons $A=208$. The results obtained with the two methods are consistent within uncertainties.

The combined b-jet cross sections are compared with NLO pQCD calculations by the POWHEG dijet tune with PYTHIA~8 fragmentation~\cite{Alioli:2010xa,Frixione:2007vw}, see Fig.~\ref{fig:bjet_xsec_model}. The measured b-jet cross section is described by the calculations within the experimental and theoretical uncertainties. The quoted theoretical uncertainties on the POWHEG data contain uncertainties obtained by changing the renormalization and factorization scales by a factor 0.5--2, variation of $\alpha_s$, and variation of the PDFs of the CT14NLO parton distribution function~\cite{Dulat:2015mca} and the EPPS16 nPDF~\cite{Eskola:2016oht} in the POWHEG calculations. The uncertainties from CT14NLO and EPPS16 were propagated according to the Hessian prescription of the authors of these parameterizations (Eq. 53 of Ref.~\cite{Eskola:2016oht}). 
The uncertainty on $\alpha_s$ was estimated 
by varying the strong coupling from 0.111 to 0.123. 

\subsection{b-jet fraction}\label{sec:bjet_Fraction}

\begin{figure}[h!]%
\centering
{\includegraphics[width=.5\textwidth]{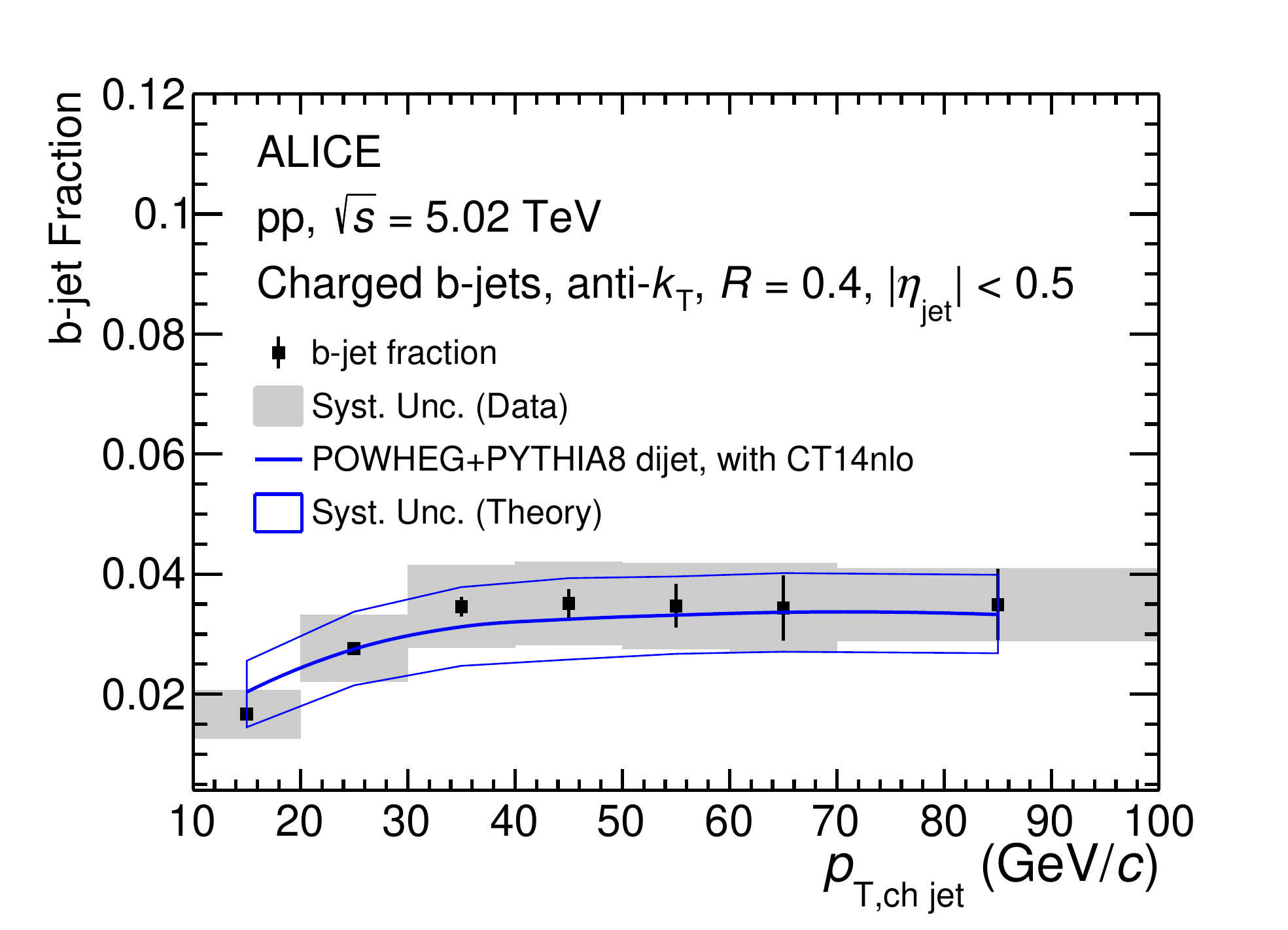}}%
{\includegraphics[width=.5\textwidth]{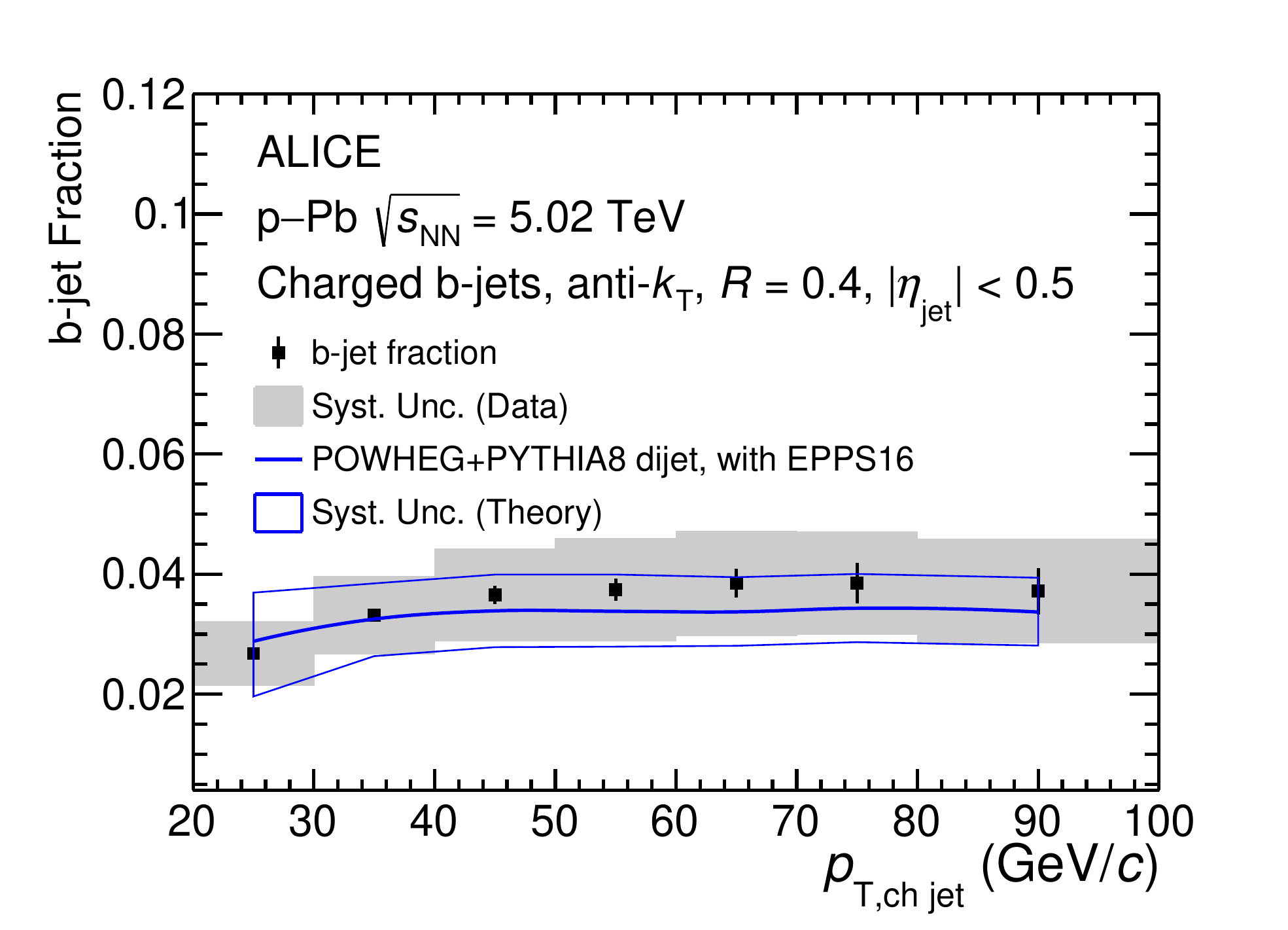}}%
\caption{The b-jet fraction in pp collisions at \five (left) and \pPb collisions at \fivenn (right), compared with POWHEG NLO pQCD calculations with PYTHIA~8 fragmentation.}
\label{fig:bfrac}
\end{figure}

Figure~\ref{fig:bfrac} shows the fraction of charged-particle b-jets among inclusive charged-particle jets in \pp and \pPb collisions. The reference \pT-differential cross sections of inclusive charged-particle jet production in \pp and \pPb were taken from Refs.~\cite{Acharya:2019tku} and~\cite{Adam:2015hoa}, respectively. 
The inclusive-jet and the b-jet measurements were obtained from different data samples, collected in different periods. Although the uncertainties corresponding to track reconstruction may be partly correlated, as a conservative approach, both the statistical and systematic uncertainties of the inclusive and b-jet cross sections were considered as uncorrelated.
The measured b-jet fractions are compared with calculations of the POWHEG dijet tune with PYTHIA~8 fragmentation~\cite{Alioli:2010xa,Frixione:2007vw}. In the \pPb case, the EPPS16 nuclear PDF set was also applied. The measured b-jet fraction is described by these calculations within uncertainties.

\subsection{The b-jet nuclear modification factor \RpPbBjet}
\label{sec:bjet_RpPb}

\begin{figure}[h!]%
\centering
{\includegraphics[width=.5\textwidth]{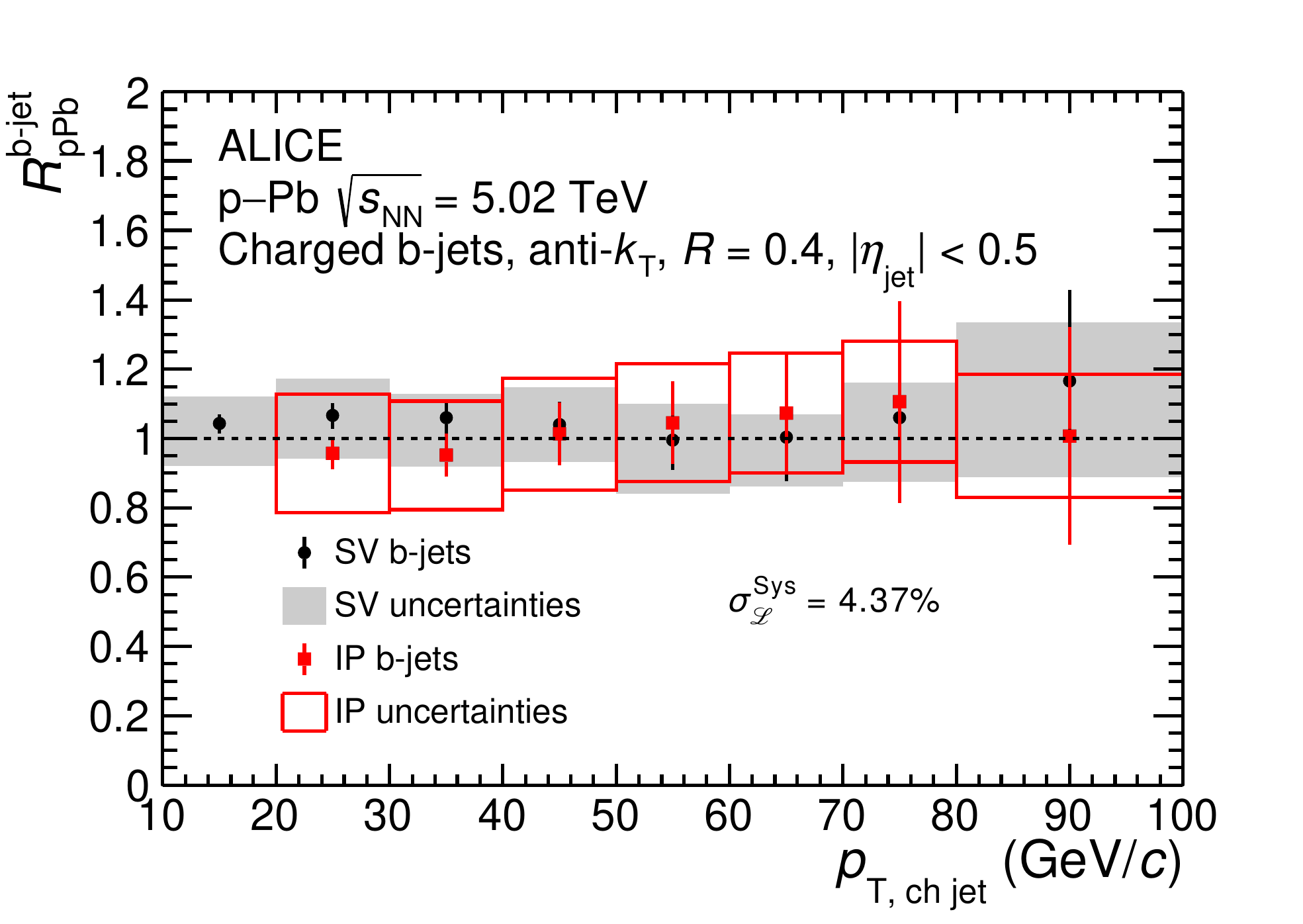}}%
{\includegraphics[width=.5\textwidth]{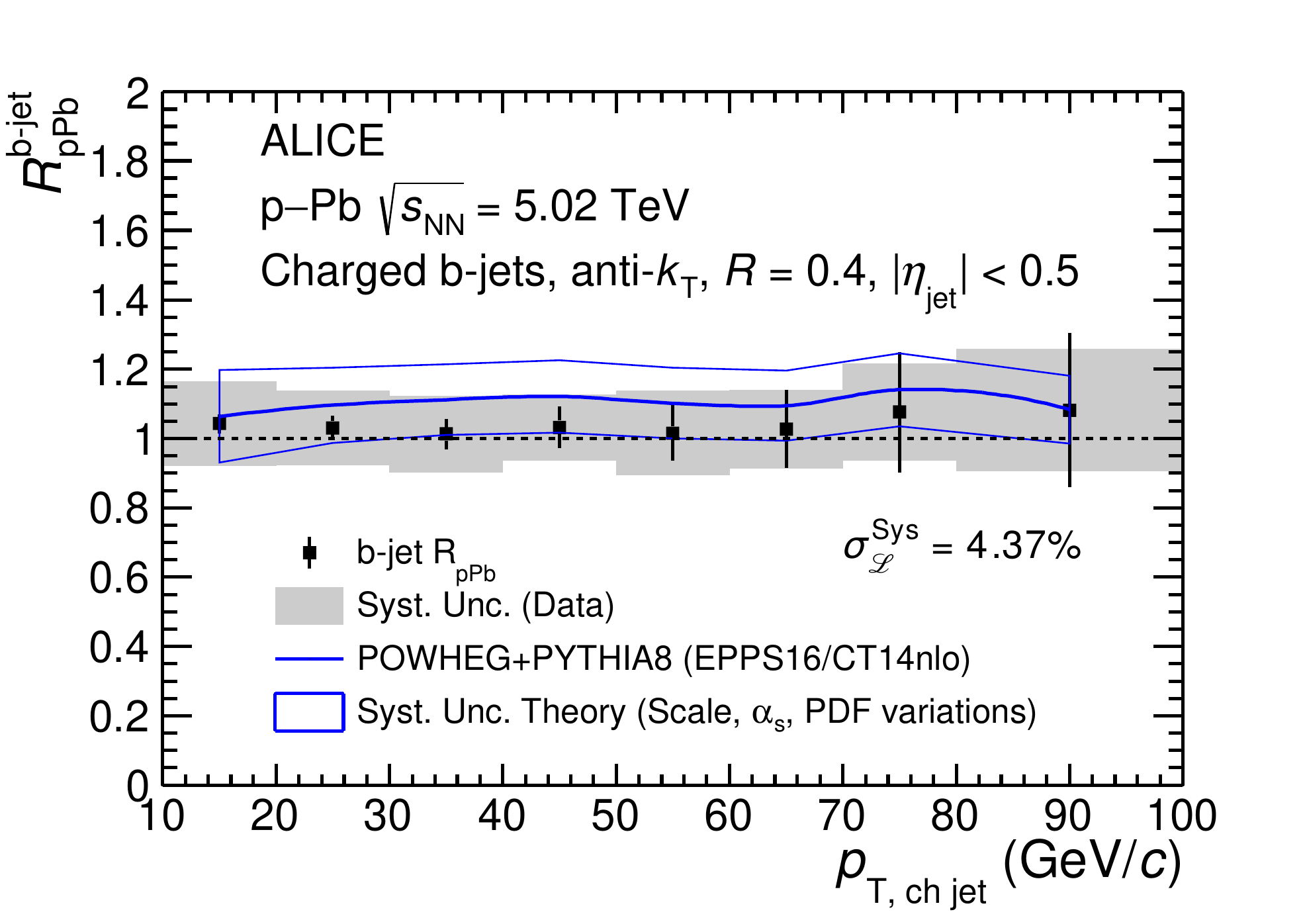}}%
\caption{Left: The nuclear modification factor \RpPbBjet of the inclusive charged-particle anti-$k_{\rm T}$ $R=0.4$ b jets as a function of \pt from the IP and SV method. Right: The nuclear modification factor \RpPbBjet obtained from combining the IP and SV method results as a function of \pTchjet compared with the calculation by the POWHEG dijet tune with the PYTHIA~8 fragmentation~\cite{Alioli:2010xa,Frixione:2007vw}.
Systematic and statistical uncertainties are shown as boxes and error bars, respectively. There is an additional normalization uncertainty of $4.37\%$ due to luminosity, which is quoted separately.}
\label{fig:bjet_RpPb}
\end{figure}

Figure~\ref{fig:bjet_RpPb} (left) shows the nuclear modification factor of charged-particle b jets obtained from the IP and SV methods. The \RpPbBjet of the two methods are consistent within uncertainties. Figure~\ref{fig:bjet_RpPb} (right) displays the combined b-jet nuclear modification factor \RpPbBjet as a function of \pTchjet, compared to the NLO pQCD, POWHEG dijet tune with PYTHIA~8 fragmentation calculation~\cite{Alioli:2010xa,Frixione:2007vw}. The NLO \RpPbBjet was estimated from the ratio of the b-jet spectra obtained with EPPS16 and CT14NLO parton distribution functions. The \RpPbBjet is consistent with unity within uncertainties, as well as with a mild modification of $\RpPbBjet\approx 1.1\pm0.1$ predicted by antishadowing in the EPPS16 nuclear PDFs, in the full $10<\pTchjet<100$\,\GeVc range of the measurement. The pQCD calculations describe the data within uncertainties. These results indicate that there are no strong nuclear matter effects present in b-jet production at midrapidity in \pPb collisions at \fivenn.

\begin{figure}[h!]%
\centering
{\includegraphics[width=.65\linewidth]{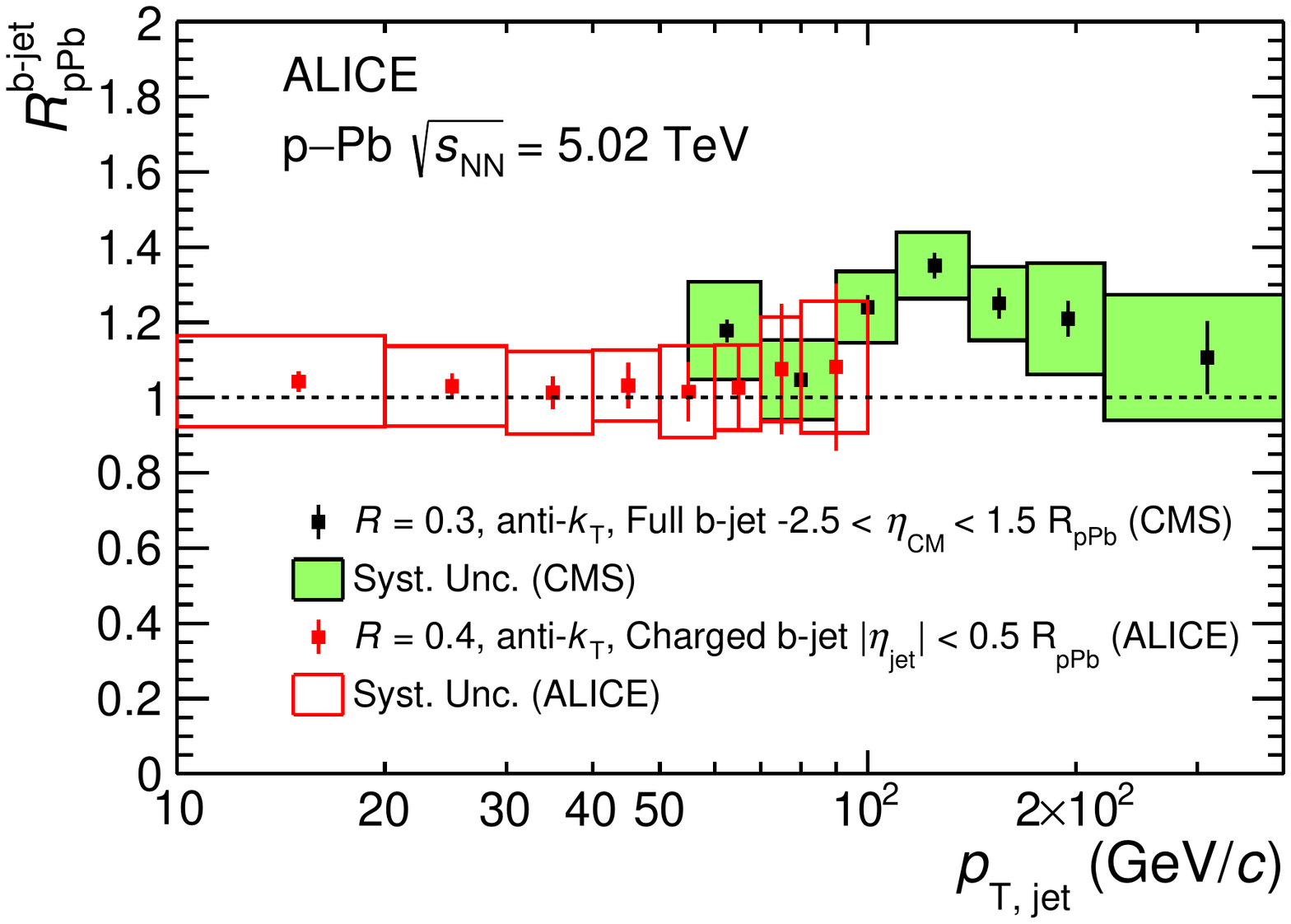}}%

\caption{The nuclear modification factor \RpPbBjet for charged-particle b jets measured by the ALICE experiment, compared with the b-jet measurement from the CMS experiment~\cite{Khachatryan:2015sva}. The CMS measurement represents $R=0.3$ fully reconstructed b jets within $-2.5<\eta_{\rm jet}<1.5$. There is an additional 22\% scaling uncertainty from the PYTHIA \pp reference on the CMS data that is not shown in the figure. The ALICE \RpPbBjet data have an additional normalization uncertainty of 
$4.37\%$.}

\label{fig:CMS_bjet_RpPb}
\end{figure}
Figure~\ref{fig:CMS_bjet_RpPb} shows the \RpPbBjet for charged-particle b jets measured by ALICE as a function of jet \pt, compared with the measurement of the CMS collaboration for full-jet b jets~\cite{Khachatryan:2015sva}.
Since the jets from CMS also include the neutral particles, the \pT scales do not compare directly. Note that there is an additional $\approx22$\% scaling uncertainty on the CMS data from the \pp reference that was computed using PYTHIA simulations.
Despite the different jet definitions and rapidity ranges used in the two measurements, the ALICE and CMS data are fully compatible in the overlap region.
A substantial nuclear modification of b-jet production by cold nuclear matter can be excluded in the whole range from $\pTchjet>10$\,\GeVc (approximately corresponding to $p_{\rm T,full jet} \gtrsim 15$\,\GeVc) up to $p_{\rm T,full jet}<400$\,\GeVc.

\begin{figure}[h!]%
\centering
{\includegraphics[width=.65\linewidth]{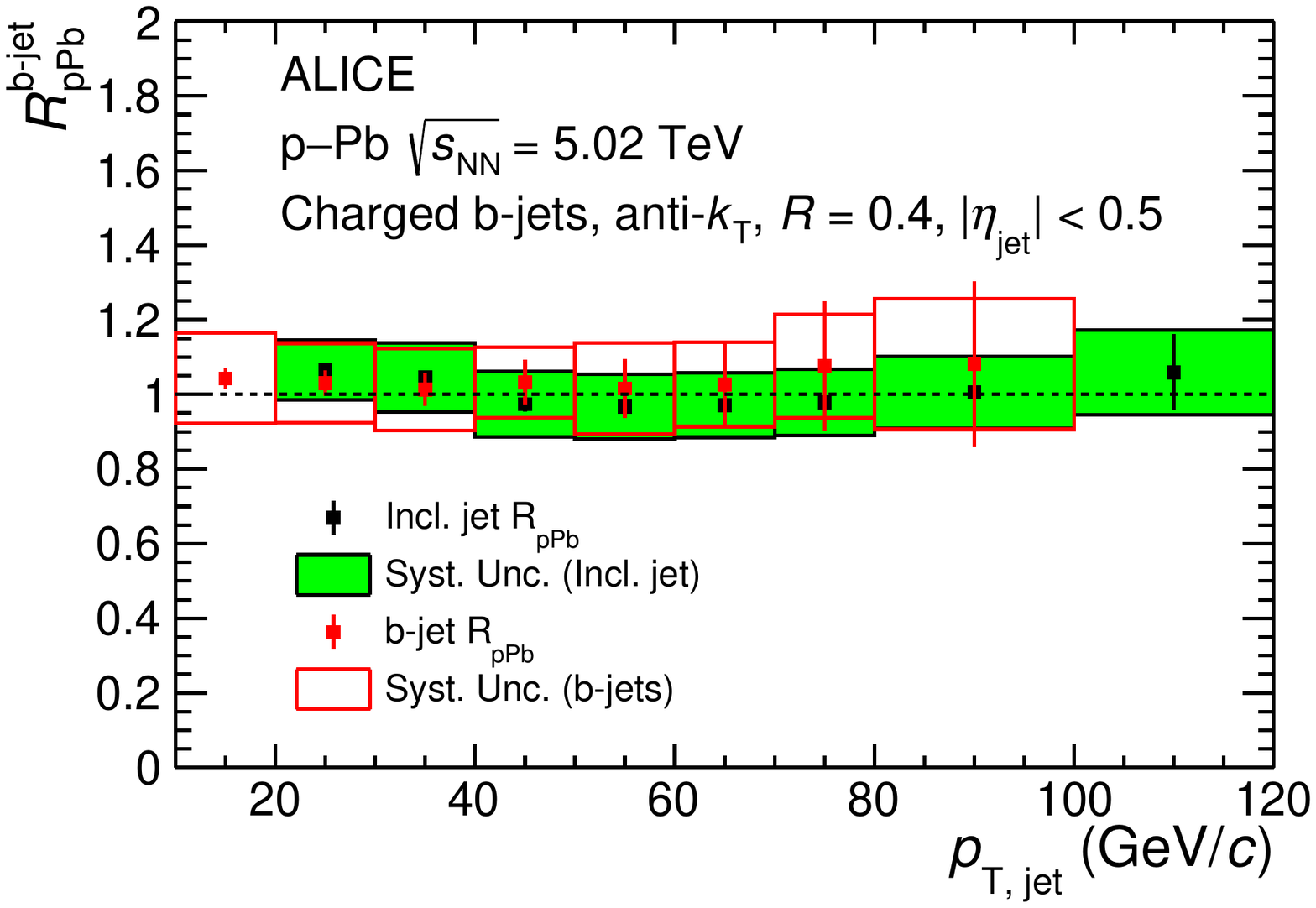}}%

\caption{The nuclear modification factor of b jets compared to that of inclusive jets from Ref.~\cite{Adam:2015hoa}.
There is a global normalization uncertainty of $4.37\%$ on the \RpPbBjet data from luminosity calculation, and $11.6\%$ on the inclusive-jet \RpPb data due to luminosity calculation and the scaling of the \pp reference.}

\label{fig:Incjet_RpPb_Comp}
\end{figure}

Fig.~\ref{fig:Incjet_RpPb_Comp} shows the nuclear modification factor of the b jets compared to that of inclusive jets from Ref.~\cite{Adam:2015hoa}. The inclusive-jet \RpPb, dominated by jets originating from light-flavor gluons and quarks, is consistent with unity as well as with \RpPbBjet. This suggests that jets in the given \pTchjet may only be subject to mild cold nuclear matter effects, regardless of the jet-initiating parton.

\section{Summary} 
\label{sec:summary}

A measurement of the \pt-differential b-jet production cross sections in \pp and \pPb collisions at 
$\sqrt{s_{\rm NN}} = 5.02$\, TeV is presented in this paper, in the transverse momentum range $10 \le \pTchjet \le 100$\,\GeVc and the central rapidity region. 
The lower \pT reach of the current measurements is unprecedented at the LHC.
The fraction of b jets compared to inclusive jets in \pp collisions are around 0.02 in the lowest $10\le \pTchjet < 20$\,\GeVc interval, saturating at about 0.03 from $\pTchjet \ge 30$\,\GeVc. There is no significant difference between the b-jet fractions measured in \pp and \pPb collisions.
The nuclear modification factor \RpPbBjet is found to be consistent with unity within the current precision, implying no strong cold nuclear effects on the b-jet production in \pPb collisions at \fivenn. The b-jet measurements are described by NLO pQCD POWHEG calculations with PYTHIA 8 fragmentation within uncertainties.

In the low jet transverse momentum range, jet energy loss by radiative and collisional mechanisms in a hot and dense medium is expected to be strongly mass dependent. The current results, which exploit the excellent tracking capabilities of ALICE and reach down to $\pTchjet=10$\,\GeVc, provide a baseline for future measurements of nuclear modification in \PbPb collisions.


\newenvironment{acknowledgement}{\relax}{\relax}
\begin{acknowledgement}
\section*{Acknowledgements}

The ALICE Collaboration would like to thank all its engineers and technicians for their invaluable contributions to the construction of the experiment and the CERN accelerator teams for the outstanding performance of the LHC complex.
The ALICE Collaboration gratefully acknowledges the resources and support provided by all Grid centres and the Worldwide LHC Computing Grid (WLCG) collaboration.
The ALICE Collaboration acknowledges the following funding agencies for their support in building and running the ALICE detector:
A. I. Alikhanyan National Science Laboratory (Yerevan Physics Institute) Foundation (ANSL), State Committee of Science and World Federation of Scientists (WFS), Armenia;
Austrian Academy of Sciences, Austrian Science Fund (FWF): [M 2467-N36] and Nationalstiftung f\"{u}r Forschung, Technologie und Entwicklung, Austria;
Ministry of Communications and High Technologies, National Nuclear Research Center, Azerbaijan;
Conselho Nacional de Desenvolvimento Cient\'{\i}fico e Tecnol\'{o}gico (CNPq), Financiadora de Estudos e Projetos (Finep), Funda\c{c}\~{a}o de Amparo \`{a} Pesquisa do Estado de S\~{a}o Paulo (FAPESP) and Universidade Federal do Rio Grande do Sul (UFRGS), Brazil;
Ministry of Education of China (MOEC) , Ministry of Science \& Technology of China (MSTC) and National Natural Science Foundation of China (NSFC), China;
Ministry of Science and Education and Croatian Science Foundation, Croatia;
Centro de Aplicaciones Tecnol\'{o}gicas y Desarrollo Nuclear (CEADEN), Cubaenerg\'{\i}a, Cuba;
Ministry of Education, Youth and Sports of the Czech Republic, Czech Republic;
The Danish Council for Independent Research | Natural Sciences, the VILLUM FONDEN and Danish National Research Foundation (DNRF), Denmark;
Helsinki Institute of Physics (HIP), Finland;
Commissariat \`{a} l'Energie Atomique (CEA) and Institut National de Physique Nucl\'{e}aire et de Physique des Particules (IN2P3) and Centre National de la Recherche Scientifique (CNRS), France;
Bundesministerium f\"{u}r Bildung und Forschung (BMBF) and GSI Helmholtzzentrum f\"{u}r Schwerionenforschung GmbH, Germany;
General Secretariat for Research and Technology, Ministry of Education, Research and Religions, Greece;
National Research, Development and Innovation Office, Hungary;
Department of Atomic Energy Government of India (DAE), Department of Science and Technology, Government of India (DST), University Grants Commission, Government of India (UGC) and Council of Scientific and Industrial Research (CSIR), India;
Indonesian Institute of Science, Indonesia;
Istituto Nazionale di Fisica Nucleare (INFN), Italy;
Japanese Ministry of Education, Culture, Sports, Science and Technology (MEXT), Japan Society for the Promotion of Science (JSPS) KAKENHI and Japanese Ministry of Education, Culture, Sports, Science and Technology (MEXT)of Applied Science (IIST), Japan;
Consejo Nacional de Ciencia (CONACYT) y Tecnolog\'{i}a, through Fondo de Cooperaci\'{o}n Internacional en Ciencia y Tecnolog\'{i}a (FONCICYT) and Direcci\'{o}n General de Asuntos del Personal Academico (DGAPA), Mexico;
Nederlandse Organisatie voor Wetenschappelijk Onderzoek (NWO), Netherlands;
The Research Council of Norway, Norway;
Commission on Science and Technology for Sustainable Development in the South (COMSATS), Pakistan;
Pontificia Universidad Cat\'{o}lica del Per\'{u}, Peru;
Ministry of Education and Science, National Science Centre and WUT ID-UB, Poland;
Korea Institute of Science and Technology Information and National Research Foundation of Korea (NRF), Republic of Korea;
Ministry of Education and Scientific Research, Institute of Atomic Physics and Ministry of Research and Innovation and Institute of Atomic Physics, Romania;
Joint Institute for Nuclear Research (JINR), Ministry of Education and Science of the Russian Federation, National Research Centre Kurchatov Institute, Russian Science Foundation and Russian Foundation for Basic Research, Russia;
Ministry of Education, Science, Research and Sport of the Slovak Republic, Slovakia;
National Research Foundation of South Africa, South Africa;
Swedish Research Council (VR) and Knut \& Alice Wallenberg Foundation (KAW), Sweden;
European Organization for Nuclear Research, Switzerland;
Suranaree University of Technology (SUT), National Science and Technology Development Agency (NSDTA) and Office of the Higher Education Commission under NRU project of Thailand, Thailand;
Turkish Energy, Nuclear and Mineral Research Agency (TENMAK), Turkey;
National Academy of  Sciences of Ukraine, Ukraine;
Science and Technology Facilities Council (STFC), United Kingdom;
National Science Foundation of the United States of America (NSF) and United States Department of Energy, Office of Nuclear Physics (DOE NP), United States of America.
In addition, individual groups and members have received support from the Lebanese National Council for Scientific Research (CNRS-L) and Lebanese University, Lebanon.

\end{acknowledgement}

\bibliographystyle{utphys}   
\bibliography{bibliography}

\newpage
\appendix

%
%

\section{The ALICE Collaboration}
\label{app:collab}
\small
\begin{flushleft} 

S.~Acharya$^{\rm 142}$, 
D.~Adamov\'{a}$^{\rm 97}$, 
A.~Adler$^{\rm 75}$, 
J.~Adolfsson$^{\rm 82}$, 
G.~Aglieri Rinella$^{\rm 34}$, 
M.~Agnello$^{\rm 30}$, 
N.~Agrawal$^{\rm 54}$, 
Z.~Ahammed$^{\rm 142}$, 
S.~Ahmad$^{\rm 16}$, 
S.U.~Ahn$^{\rm 77}$
% ...

\section*{Affiliation notes}

$^{\rm I}$ Deceased\\
$^{\rm II}$ Also at: Italian National Agency for New Technologies, Energy and Sustainable Economic Development (ENEA), Bologna, Italy\\
$^{\rm III}$ Also at: Dipartimento DET del Politecnico di Torino, Turin, Italy\\
$^{\rm IV}$ Also at: M.V. Lomonosov Moscow State University, D.V. Skobeltsyn Institute of Nuclear, Physics, Moscow, Russia\\
$^{\rm V}$ Also at: Department of Applied Physics, Aligarh Muslim University, Aligarh, India
\\
$^{\rm VI}$ Also at: Institute of Theoretical Physics, University of Wroclaw, Poland\\
$^{\rm VII}$ Also at: University of Kansas, Lawrence, Kansas, United States\\

\section*{Collaboration Institutes}

$^{1}$ A.I. Alikhanyan National Science Laboratory (Yerevan Physics Institute) Foundation, Yerevan, Armenia\\
$^{2}$ AGH University of Science and Technology, Cracow, Poland\\
$^{3}$ Bogolyubov Institute for Theoretical Physics, National Academy of Sciences of Ukraine, Kiev, Ukraine\\
$^{4}$ Bose Institute, Department of Physics  and Centre for Astroparticle Physics and Space Science (CAPSS), Kolkata, India\\
$^{5}$ Budker Institute for Nuclear Physics, Novosibirsk, Russia\\
$^{6}$ California Polytechnic State University, San Luis Obispo, California, United States\\
$^{7}$ Central China Normal University, Wuhan, China\\
$^{8}$ Centro de Aplicaciones Tecnol\'{o}gicas y Desarrollo Nuclear (CEADEN), Havana, Cuba\\
$^{9}$ Centro de Investigaci\'{o}n y de Estudios Avanzados (CINVESTAV), Mexico City and M\'{e}rida, Mexico\\
$^{10}$ Chicago State University, Chicago, Illinois, United States\\
$^{11}$ China Institute of Atomic Energy, Beijing, China\\
$^{12}$ Chungbuk National University, Cheongju, Republic of Korea\\
$^{13}$ Comenius University Bratislava, Faculty of Mathematics, Physics and Informatics, Bratislava, Slovakia\\
$^{14}$ COMSATS University Islamabad, Islamabad, Pakistan\\
$^{15}$ Creighton University, Omaha, Nebraska, United States\\
$^{16}$ Department of Physics, Aligarh Muslim University, Aligarh, India\\
$^{17}$ Department of Physics, Pusan National University, Pusan, Republic of Korea\\
$^{18}$ Department of Physics, Sejong University, Seoul, Republic of Korea\\
$^{19}$ Department of Physics, University of California, Berkeley, California, United States\\
$^{20}$ Department of Physics, University of Oslo, Oslo, Norway\\
$^{21}$ Department of Physics and Technology, University of Bergen, Bergen, Norway\\
$^{22}$ Dipartimento di Fisica dell'Universit\`{a} and Sezione INFN, Cagliari, Italy\\
$^{23}$ Dipartimento di Fisica dell'Universit\`{a} and Sezione INFN, Trieste, Italy\\
$^{24}$ Dipartimento di Fisica dell'Universit\`{a} and Sezione INFN, Turin, Italy\\
$^{25}$ Dipartimento di Fisica e Astronomia dell'Universit\`{a} and Sezione INFN, Bologna, Italy\\
$^{26}$ Dipartimento di Fisica e Astronomia dell'Universit\`{a} and Sezione INFN, Catania, Italy\\
$^{27}$ Dipartimento di Fisica e Astronomia dell'Universit\`{a} and Sezione INFN, Padova, Italy\\
$^{28}$ Dipartimento di Fisica e Nucleare e Teorica, Universit\`{a} di Pavia, Pavia, Italy\\
$^{29}$ Dipartimento di Fisica `E.R.~Caianiello' dell'Universit\`{a} and Gruppo Collegato INFN, Salerno, Italy\\
$^{30}$ Dipartimento DISAT del Politecnico and Sezione INFN, Turin, Italy\\
$^{31}$ Dipartimento di Scienze e Innovazione Tecnologica dell'Universit\`{a} del Piemonte Orientale and INFN Sezione di Torino, Alessandria, Italy\\
$^{32}$ Dipartimento di Scienze MIFT, Universit\`{a} di Messina, Messina, Italy\\
$^{33}$ Dipartimento Interateneo di Fisica `M.~Merlin' and Sezione INFN, Bari, Italy\\
$^{34}$ European Organization for Nuclear Research (CERN), Geneva, Switzerland\\
$^{35}$ Faculty of Electrical Engineering, Mechanical Engineering and Naval Architecture, University of Split, Split, Croatia\\
$^{36}$ Faculty of Engineering and Science, Western Norway University of Applied Sciences, Bergen, Norway\\
$^{37}$ Faculty of Nuclear Sciences and Physical Engineering, Czech Technical University in Prague, Prague, Czech Republic\\
$^{38}$ Faculty of Science, P.J.~\v{S}af\'{a}rik University, Ko\v{s}ice, Slovakia\\
$^{39}$ Frankfurt Institute for Advanced Studies, Johann Wolfgang Goethe-Universit\"{a}t Frankfurt, Frankfurt, Germany\\
$^{40}$ Fudan University, Shanghai, China\\
$^{41}$ Gangneung-Wonju National University, Gangneung, Republic of Korea\\
$^{42}$ Gauhati University, Department of Physics, Guwahati, India\\
$^{43}$ Helmholtz-Institut f\"{u}r Strahlen- und Kernphysik, Rheinische Friedrich-Wilhelms-Universit\"{a}t Bonn, Bonn, Germany\\
$^{44}$ Helsinki Institute of Physics (HIP), Helsinki, Finland\\
$^{45}$ High Energy Physics Group,  Universidad Aut\'{o}noma de Puebla, Puebla, Mexico\\
$^{46}$ Hiroshima University, Hiroshima, Japan\\
$^{47}$ Hochschule Worms, Zentrum  f\"{u}r Technologietransfer und Telekommunikation (ZTT), Worms, Germany\\
$^{48}$ Horia Hulubei National Institute of Physics and Nuclear Engineering, Bucharest, Romania\\
$^{49}$ Indian Institute of Technology Bombay (IIT), Mumbai, India\\
$^{50}$ Indian Institute of Technology Indore, Indore, India\\
$^{51}$ Indonesian Institute of Sciences, Jakarta, Indonesia\\
$^{52}$ INFN, Laboratori Nazionali di Frascati, Frascati, Italy\\
$^{53}$ INFN, Sezione di Bari, Bari, Italy\\
$^{54}$ INFN, Sezione di Bologna, Bologna, Italy\\
$^{55}$ INFN, Sezione di Cagliari, Cagliari, Italy\\
$^{56}$ INFN, Sezione di Catania, Catania, Italy\\
$^{57}$ INFN, Sezione di Padova, Padova, Italy\\
$^{58}$ INFN, Sezione di Pavia, Pavia, Italy\\
$^{59}$ INFN, Sezione di Roma, Rome, Italy\\
$^{60}$ INFN, Sezione di Torino, Turin, Italy\\
$^{61}$ INFN, Sezione di Trieste, Trieste, Italy\\
$^{62}$ Inha University, Incheon, Republic of Korea\\
$^{63}$ Institute for Gravitational and Subatomic Physics (GRASP), Utrecht University/Nikhef, Utrecht, Netherlands\\
$^{64}$ Institute for Nuclear Research, Academy of Sciences, Moscow, Russia\\
$^{65}$ Institute of Experimental Physics, Slovak Academy of Sciences, Ko\v{s}ice, Slovakia\\
$^{66}$ Institute of Physics, Homi Bhabha National Institute, Bhubaneswar, India\\
$^{67}$ Institute of Physics of the Czech Academy of Sciences, Prague, Czech Republic\\
$^{68}$ Institute of Space Science (ISS), Bucharest, Romania\\
$^{69}$ Institut f\"{u}r Kernphysik, Johann Wolfgang Goethe-Universit\"{a}t Frankfurt, Frankfurt, Germany\\
$^{70}$ Instituto de Ciencias Nucleares, Universidad Nacional Aut\'{o}noma de M\'{e}xico, Mexico City, Mexico\\
$^{71}$ Instituto de F\'{i}sica, Universidade Federal do Rio Grande do Sul (UFRGS), Porto Alegre, Brazil\\
$^{72}$ Instituto de F\'{\i}sica, Universidad Nacional Aut\'{o}noma de M\'{e}xico, Mexico City, Mexico\\
$^{73}$ iThemba LABS, National Research Foundation, Somerset West, South Africa\\
$^{74}$ Jeonbuk National University, Jeonju, Republic of Korea\\
$^{75}$ Johann-Wolfgang-Goethe Universit\"{a}t Frankfurt Institut f\"{u}r Informatik, Fachbereich Informatik und Mathematik, Frankfurt, Germany\\
$^{76}$ Joint Institute for Nuclear Research (JINR), Dubna, Russia\\
$^{77}$ Korea Institute of Science and Technology Information, Daejeon, Republic of Korea\\
$^{78}$ KTO Karatay University, Konya, Turkey\\
$^{79}$ Laboratoire de Physique des 2 Infinis, Ir\`{e}ne Joliot-Curie, Orsay, France\\
$^{80}$ Laboratoire de Physique Subatomique et de Cosmologie, Universit\'{e} Grenoble-Alpes, CNRS-IN2P3, Grenoble, France\\
$^{81}$ Lawrence Berkeley National Laboratory, Berkeley, California, United States\\
$^{82}$ Lund University Department of Physics, Division of Particle Physics, Lund, Sweden\\
$^{83}$ Moscow Institute for Physics and Technology, Moscow, Russia\\
$^{84}$ Nagasaki Institute of Applied Science, Nagasaki, Japan\\
$^{85}$ Nara Women{'}s University (NWU), Nara, Japan\\
$^{86}$ National and Kapodistrian University of Athens, School of Science, Department of Physics , Athens, Greece\\
$^{87}$ National Centre for Nuclear Research, Warsaw, Poland\\
$^{88}$ National Institute of Science Education and Research, Homi Bhabha National Institute, Jatni, India\\
$^{89}$ National Nuclear Research Center, Baku, Azerbaijan\\
$^{90}$ National Research Centre Kurchatov Institute, Moscow, Russia\\
$^{91}$ Niels Bohr Institute, University of Copenhagen, Copenhagen, Denmark\\
$^{92}$ Nikhef, National institute for subatomic physics, Amsterdam, Netherlands\\
$^{93}$ NRC Kurchatov Institute IHEP, Protvino, Russia\\
$^{94}$ NRC \guillemotleft Kurchatov\guillemotright  Institute - ITEP, Moscow, Russia\\
$^{95}$ NRNU Moscow Engineering Physics Institute, Moscow, Russia\\
$^{96}$ Nuclear Physics Group, STFC Daresbury Laboratory, Daresbury, United Kingdom\\
$^{97}$ Nuclear Physics Institute of the Czech Academy of Sciences, \v{R}e\v{z} u Prahy, Czech Republic\\
$^{98}$ Oak Ridge National Laboratory, Oak Ridge, Tennessee, United States\\
$^{99}$ Ohio State University, Columbus, Ohio, United States\\
$^{100}$ Petersburg Nuclear Physics Institute, Gatchina, Russia\\
$^{101}$ Physics department, Faculty of science, University of Zagreb, Zagreb, Croatia\\
$^{102}$ Physics Department, Panjab University, Chandigarh, India\\
$^{103}$ Physics Department, University of Jammu, Jammu, India\\
$^{104}$ Physics Department, University of Rajasthan, Jaipur, India\\
$^{105}$ Physikalisches Institut, Eberhard-Karls-Universit\"{a}t T\"{u}bingen, T\"{u}bingen, Germany\\
$^{106}$ Physikalisches Institut, Ruprecht-Karls-Universit\"{a}t Heidelberg, Heidelberg, Germany\\
$^{107}$ Physik Department, Technische Universit\"{a}t M\"{u}nchen, Munich, Germany\\
$^{108}$ Politecnico di Bari and Sezione INFN, Bari, Italy\\
$^{109}$ Research Division and ExtreMe Matter Institute EMMI, GSI Helmholtzzentrum f\"ur Schwerionenforschung GmbH, Darmstadt, Germany\\
$^{110}$ Russian Federal Nuclear Center (VNIIEF), Sarov, Russia\\
$^{111}$ Saha Institute of Nuclear Physics, Homi Bhabha National Institute, Kolkata, India\\
$^{112}$ School of Physics and Astronomy, University of Birmingham, Birmingham, United Kingdom\\
$^{113}$ Secci\'{o}n F\'{\i}sica, Departamento de Ciencias, Pontificia Universidad Cat\'{o}lica del Per\'{u}, Lima, Peru\\
$^{114}$ St. Petersburg State University, St. Petersburg, Russia\\
$^{115}$ Stefan Meyer Institut f\"{u}r Subatomare Physik (SMI), Vienna, Austria\\
$^{116}$ SUBATECH, IMT Atlantique, Universit\'{e} de Nantes, CNRS-IN2P3, Nantes, France\\
$^{117}$ Suranaree University of Technology, Nakhon Ratchasima, Thailand\\
$^{118}$ Technical University of Ko\v{s}ice, Ko\v{s}ice, Slovakia\\
$^{119}$ The Henryk Niewodniczanski Institute of Nuclear Physics, Polish Academy of Sciences, Cracow, Poland\\
$^{120}$ The University of Texas at Austin, Austin, Texas, United States\\
$^{121}$ Universidad Aut\'{o}noma de Sinaloa, Culiac\'{a}n, Mexico\\
$^{122}$ Universidade de S\~{a}o Paulo (USP), S\~{a}o Paulo, Brazil\\
$^{123}$ Universidade Estadual de Campinas (UNICAMP), Campinas, Brazil\\
$^{124}$ Universidade Federal do ABC, Santo Andre, Brazil\\
$^{125}$ University of Cape Town, Cape Town, South Africa\\
$^{126}$ University of Houston, Houston, Texas, United States\\
$^{127}$ University of Jyv\"{a}skyl\"{a}, Jyv\"{a}skyl\"{a}, Finland\\
$^{128}$ University of Kansas, Lawrence, Kansas, United States\\
$^{129}$ University of Liverpool, Liverpool, United Kingdom\\
$^{130}$ University of Science and Technology of China, Hefei, China\\
$^{131}$ University of South-Eastern Norway, Tonsberg, Norway\\
$^{132}$ University of Tennessee, Knoxville, Tennessee, United States\\
$^{133}$ University of the Witwatersrand, Johannesburg, South Africa\\
$^{134}$ University of Tokyo, Tokyo, Japan\\
$^{135}$ University of Tsukuba, Tsukuba, Japan\\
$^{136}$ Universit\'{e} Clermont Auvergne, CNRS/IN2P3, LPC, Clermont-Ferrand, France\\
$^{137}$ Universit\'{e} de Lyon, CNRS/IN2P3, Institut de Physique des 2 Infinis de Lyon, Lyon, France\\
$^{138}$ Universit\'{e} de Strasbourg, CNRS, IPHC UMR 7178, F-67000 Strasbourg, France, Strasbourg, France\\
$^{139}$ Universit\'{e} Paris-Saclay Centre d'Etudes de Saclay (CEA), IRFU, D\'{e}partment de Physique Nucl\'{e}aire (DPhN), Saclay, France\\
$^{140}$ Universit\`{a} degli Studi di Foggia, Foggia, Italy\\
$^{141}$ Universit\`{a} di Brescia, Brescia, Italy\\
$^{142}$ Variable Energy Cyclotron Centre, Homi Bhabha National Institute, Kolkata, India\\
$^{143}$ Warsaw University of Technology, Warsaw, Poland\\
$^{144}$ Wayne State University, Detroit, Michigan, United States\\
$^{145}$ Westf\"{a}lische Wilhelms-Universit\"{a}t M\"{u}nster, Institut f\"{u}r Kernphysik, M\"{u}nster, Germany\\
$^{146}$ Wigner Research Centre for Physics, Budapest, Hungary\\
$^{147}$ Yale University, New Haven, Connecticut, United States\\
$^{148}$ Yonsei University, Seoul, Republic of Korea\\

\end{flushleft} 
  
\end{document}